\documentclass[preprint,12pt]{elsarticle}

\usepackage{amssymb}
\usepackage{tikz}
\usepackage{subfigure}
\usepackage{amsmath}

\begin{document}

\begin{frontmatter}



\title{Knowledge-embedded meta-learning model for \\
lift coefficient prediction of airfoils}


\author[inst1]{Hairun Xie}
\author[inst2]{Jing Wang\corref{cor1}}
\ead{wangjinger@sjtu.edu.cn}
\cortext[cor1]{Corresponding author.}

\author[inst1]{Miao Zhang}

\affiliation[inst1]{organization={Shanghai Aircraft Design and Research Institute},
            city={Shanghai},
            postcode={200436}, 
            country={China}}

\affiliation[inst2]{organization={School of Aeronautics and Astronautics, Shanghai Jiao Tong University},
            city={Shanghai},
            postcode={200240}, 
            country={China}}

\begin{abstract}
Aerodynamic performance evaluation is an important part of the aircraft aerodynamic design optimization process; 
however, traditional methods are costly and time-consuming. 
Despite the fact that various machine learning methods can achieve high accuracy, their application in engineering is still difficult due to their poor generalization performance and "black box" nature.
In this paper, a knowledge-embedded meta learning model, which fully integrates data with the theoretical knowledge of the lift curve, is developed to obtain the lift coefficients of an arbitrary supercritical airfoil under various angle of attacks. 
In the proposed model, a primary network is responsible for representing the relationship between the lift and angle of attack, while the geometry information is encoded into a hyper network to predict the unknown parameters involved in the primary network.
Specifically, three models with different architectures are trained to provide various interpretations.
Compared to the ordinary neural network, our proposed model can exhibit better generalization capability with competitive prediction accuracy.
Afterward, interpretable analysis is performed based on the Integrated Gradients and Saliency methods.
Results show that the proposed model can tend to assess the influence of airfoil geometry to the physical characteristics.
Furthermore, the exceptions and shortcomings caused by the proposed model are analysed and discussed in detail.
\end{abstract}




\end{frontmatter}


\section{Introduction}

Aerodynamic performances of an aircraft have a significant impact on its economy, safety, comfort, and environmental friendliness.
Computational Fluid Dynamics (CFD) is the most widely used method for the evaluation of aerodynamic performances, which has played a crucial role in the development of aircraft over the past decades~\cite{johnson_thirty_2005}. 
Conventionally, CFD simulations are applied to provide the detailed flow fields, then the aerodynamic performances are calculated in the form of integral quantities.
However, the simulation is unbearably time-consuming and massive iterations of simulations are required to during the optimization design process. 
Therefore, effective and accurate access to obtain the aerodynamic performances is crucial in improving the quality of airfoils, wings and aircraft.

Data-driven surrogate models have been applied to predict the aerodynamic performances in practical engineering problems.
Traditional models, such as response surface model, radical basis function, Kriging model, support vector machine have been widely used. 
However, they usually perform poor for high-dimensional, multi-scale and nonlinear problems.
In 1990s, neural network has been applied to realize the optimization of rotor blade~\cite{lamarsh_ii_aerodynamic_1992} and multi-element airfoil~\cite{greenman_high-lift_1999}.
Research in recent years is still based on this basic idea with larger data sets, more AI computing power, and more advanced deep learning algorithms, which can achieve more accurate results over a larger range of applicability. 
Artificial neural network (ANN) has been employed in many studies~\cite{secco_artificial_2017, bouhlel_scalable_2020,li_data-based_2021} and achieved better accuracy than traditional models.
Besides, there are some researchers applied convolutional neural networks (CNN) architecture to predict aerodynamic coefficients through transforming the airfoil shape to an image~\cite{chen2020multiple, zhang2018application}.
Peng et al.\cite{peng_learning_2022} presented an ANN that can learn physical laws and aerodynamic equation while providing accurate lift coefficient predictions.
More literature can be found in \cite{li_machine_2022}, which provides an overview of the wide application of machine learning methods in aerodynamic shape optimization. 
However, the generalization performance of these surrogate models is insufficient, and the fact that they are essentially black-box makes the designers less confident in the results.
This may primarily be attributed to the complex relationship between the geometry and aerodynamic performances, which are connected through the high-dimensional flow fields described by the Navier-Stokes(N-S) equations.

In recent years, with the widespread success of deep learning and computer vision techniques, many researchers have been motivated to predict the detailed flow fields over different geometries~\cite{xiaoxiao_guo_convolutional_2016, thuerey_deep_2020, wu_deep_2020,donglin_chen_flowgan_2020,duru_cnnfoil_2021,  chen_towards_2023}.
Accordingly, the aerodynamic performances can be integrated based on the predicted flow fields.
Sekar et al.~\cite{sekar2019fast} designed a CNN and a multilayer perceptron (MLP) model to predict the incompressible laminar steady flow field over airfoils with extracted geometrical parameters, Reynolds number, and angle of attack.
They found that the lift coefficients calculated from the predicted flow fields show a good match with the CFD results and the drag coefficients show a slight mismatch with the CFD results.
Wang et al.~\cite{wang_flow_2021} proposed a Variational AutoEncoder to predict the flow fields around supercritical airfoils at different flow conditions, and the maximum relative error of drag and moment coefficient obtained from the predicted flow fields are less than 5\% and 2\%, respectively. 
The aerodynamic performances based on the flow field prediction models may exhibit better generalization ability due to the substitution of the solution of the N-S equations consistent for various geometries.
However, the prediction accuracy is characterized by the fitting ability of the model for predicting the flow fields, 
which relies heavily on training data and might produce unrealistic or inferior predictions that deviate a lot from the CFD result.
Specifically, it is challenging for the network to characterize accurately the regions with dramatic changes, such as the surface pressure used to calculate aerodynamic performances, which may lead to a relatively large error.

The strategy to improve the performance of a data-driven model lies in the enhancement of its input information, i.e., richer data and more essential prior information or domain knowledge. 
Given that obtaining data is expensive and time-consuming in our case, domain knowledge and prior information have emerged as a promising alternative.
By using domain knowledge and data information at the same time, the model can exert strong ﬁtting ability of the data-driven model and possess the stability of the expert system.
In this way, the poor generalization performance and uninterpretability of the data-driven model can be overcome, which making it possible to achieve AI trustworthiness in the future of aviation\cite{roadmap2020human,TowardAirworthiness}.
In this study, we aim to develop a knowledge-embedded meta learning model to obtain the lift coefficients of an arbitrary supercritical airfoil under various angle of attacks, which can fully integrate data with the theoretical knowledge of the lift curve. 
In the proposed model, a primary network is responsible for representing the relationship between the lift and angle of attack, while the geometry information is encoded into a hyper network to predict the unknown parameters involved in the primary network.
With this architecture, theoretical knowledge of airfoil lift is embedded in the nueral network, allowing the model to have strong generalization performance at a high accuracy. 
The interpretable analysis is further conducted so as to assess the influence of airfoil geometry to the physical characteristics.

The details of this paper are organized as follows. Section 2 introduces the basic knowledge and detailed method involved in our study. 
In Section 3, aerodynamic coefficient predicting performance is shown and the interpretability of the model are discussed. 
Concluding remarks and discussions are described in Section 4.

\section{Methodology}
In this section, we first discuss the classical theoretical knowledge about the lift, then introduce the neural network model architecture, and finally describe the detailed dataset involved in our study.

\subsection{Theoretical Knowledge of the Lift}\label{sec:knowledge}
Thin-airfoil theory is one of the earliest airfoil lift analysis theories developed in 1920s, which played an important role in early aviation practice \cite{drela2014flight,houghton_aerodynamics_2017}. 
The thin airfoil theory simulates the aerodynamic properties of an airfoil with vortex sheets.
The vortex sheet consists of a continuous vortex distribution along the chord of the airfoil.
The relationship between $C_L$ and AoA, derived from the  analytical representation of vortex distribution, is expressed as follows.
\begin{equation}\label{eq:1}
C_L=\left(C_{L_0}\right)+\frac{dC_L}{d\alpha}\alpha=\pi\left(A_1-2A_0\right)+2\pi\alpha
\end{equation}
Where $A_0$ and $A_1$ are the Fourier series coefficients for the camber line's slope,  $C_{L0}$ is the lift coefficient at zero AoA.  The slope of the lift curve indicates how rapidly lift changes with AoA, and it takes the same value $2\pi$ for all thin airfoils independent of distributions of thickness and camber. 
This theory provides the basic explanation for how the lift is produced, and gives the theoretical maximum lift slope value.
However, it is restricted to thin airfoils with small camber at small AoA, and only applicable to steady flow of ideal incompressible fluid. 

For compressible flow, there is a correction relationship based on Mach number:
\begin{equation}\label{eq:2}
C_L=\frac{C_{Li}}{\sqrt{1-M_\infty^2}\ }
\end{equation}
Where $C_{Li}$ is the incompressible lift coefficient calculated by  Eq.\eqref{eq:1} and $M_\infty$ is the free-stream Mach number.
However, the effects of more complex factors, such as thickness and viscosity, are still ignored.

\begin{figure}[htpb]
	\centering
	\tikzset{every picture/.style={line width=0.75pt}} 

\begin{tikzpicture}[x=0.75pt,y=0.75pt,yscale=-1,xscale=1,scale=1.2]

\draw  [draw opacity=0][fill={rgb, 255:red, 66; green, 211; blue, 33 }  ,fill opacity=0.1 ] (168,40.5) -- (266.8,40.5) -- (266.8,234.5) -- (168,234.5) -- cycle ;
\draw  [draw opacity=0][fill={rgb, 255:red, 255; green, 0; blue, 0 }  ,fill opacity=0.1 ] (266.8,40.5) -- (389.06,40.5) -- (389.06,234.5) -- (266.8,234.5) -- cycle ;
\draw    (168.73,210) -- (405.73,210) ;
\draw [shift={(408.73,210)}, rotate = 180] [fill={rgb, 255:red, 0; green, 0; blue, 0 }  ][line width=0.08]  [draw opacity=0] (8.93,-4.29) -- (0,0) -- (8.93,4.29) -- cycle    ;
\draw    (232.8,236.5) -- (232.8,22.76) ;
\draw [shift={(232.8,19.76)}, rotate = 90] [fill={rgb, 255:red, 0; green, 0; blue, 0 }  ][line width=0.08]  [draw opacity=0] (8.93,-4.29) -- (0,0) -- (8.93,4.29) -- cycle    ;
\draw [color={rgb, 255:red, 0; green, 0; blue, 255 }  ,draw opacity=1 ][line width=2.25]    (173.8,236.76) .. controls (214.8,181.76) and (253.73,133.5) .. (282.73,90.5) .. controls (311.73,47.5) and (360.4,43.76) .. (390.4,43.76) ;
\draw [line width=1.5]  [dash pattern={on 1.5pt off 2.25pt on 3.75pt off 3pt}]  (327.73,34.5) -- (173.8,236.76) ;
\draw   (170,238) .. controls (170,242.67) and (172.33,245) .. (177,245) -- (208.58,245) .. controls (215.25,245) and (218.58,247.33) .. (218.58,252) .. controls (218.58,247.33) and (221.91,245) .. (228.58,245)(225.58,245) -- (260.16,245) .. controls (264.83,245) and (267.16,242.67) .. (267.16,238) ;
\draw   (267.3,238) .. controls (267.3,242.67) and (269.63,245) .. (274.3,245) -- (317.73,245) .. controls (324.4,245) and (327.73,247.33) .. (327.73,252) .. controls (327.73,247.33) and (331.06,245) .. (337.73,245)(334.73,245) -- (381.16,245) .. controls (385.83,245) and (388.16,242.67) .. (388.16,238) ;
\draw    (197.69,158.5) -- (226.16,158.5) ;
\draw [shift={(229.16,158.5)}, rotate = 180] [fill={rgb, 255:red, 0; green, 0; blue, 0 }  ][line width=0.08]  [draw opacity=0] (10.72,-5.15) -- (0,0) -- (10.72,5.15) -- (7.12,0) -- cycle    ;
\draw    (267.16,147.22) -- (267.16,117.22) ;
\draw [shift={(267.16,114.22)}, rotate = 90] [fill={rgb, 255:red, 0; green, 0; blue, 0 }  ][line width=0.08]  [draw opacity=0] (10.72,-5.15) -- (0,0) -- (10.72,5.15) -- (7.12,0) -- cycle    ;

\draw (397,215.67) node [anchor=north west][inner sep=0.75pt]   [align=left] {$\displaystyle \alpha $};
\draw (204,15.67) node [anchor=north west][inner sep=0.75pt]   [align=left] {$\displaystyle \mathit{C_{L}}$};
\draw (168.33,149) node [anchor=north west][inner sep=0.75pt]   [align=left] {$\displaystyle C_{L0}$};
\draw (255,150.33) node [anchor=north west][inner sep=0.75pt]   [align=left] {$\displaystyle \alpha _{crit}$};
\draw (199.36,255.6) node [anchor=north west][inner sep=0.75pt]   [align=left] {{\fontfamily{ptm}\selectfont Linear}};
\draw (294.69,256.33) node [anchor=north west][inner sep=0.75pt]   [align=left] {{\fontfamily{ptm}\selectfont Non-Linear}};
\draw (321,12.33) node [anchor=north west][inner sep=0.75pt]   [align=left] {$\displaystyle C_{L0}^{'} \alpha +C_{L0}$};

\end{tikzpicture}
	\caption{A typical $C_L-\alpha$ curve in practice.}
	\label{fig:fig1}
\end{figure}
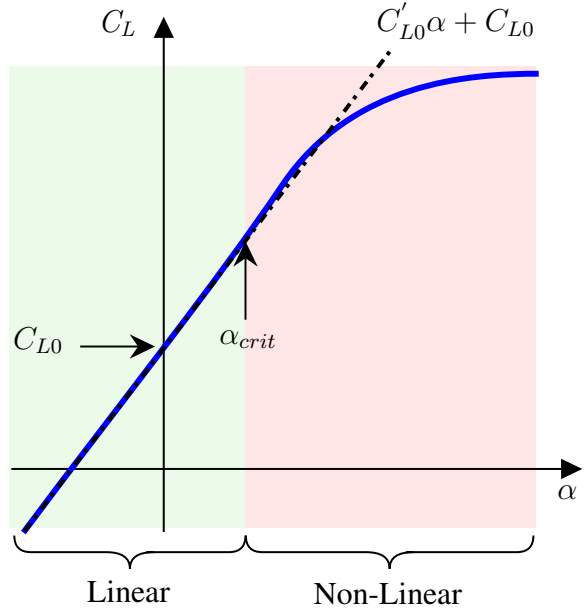

Fig.~\ref{fig:fig1} shows a typical $C_L-\alpha$ curve in practice.
When AoA is small, the curve conforms to the linear assumption relatively well.
With the increase of AoA, the intensity of the shock wave tends to  increase and attached flow appears to separate. 
When AoA increases to a certain extent, shock wave and flow separation are strong enough to interact with each other, resulting in buffet or stall.
Such flow phenomena have strong nonlinear characteristics and can cause the lift curve to deviate from the linear assumption. 
As a result, the actual lift curve can be divided into a linear portion that conforms to the linear assumption and a nonlinear portion that deviates from the linear assumption.
The actual lift curve can be described by the following physical parameters:
\begin{itemize}
\item[1)] $C_{L0}$: value of $C_L$ when $AoA=0^\circ$. 
It is 0 for symmetrical airfoils, and is less than 0 for under-cambered airfoils. 
For highly cambered airfoils, it can be greater than 1.0. 
\item[2)] $C_{L0}^\prime$: Slope of the lift curve in the linear portion.  
For the incompressible flow, theoretical values can be inferred directly. 
While for the compressible flow, corrections or more complicated relations should be derived.
\item[3)]  $\alpha_{crit}$: The transition point from the linear portion to the nonlinear portion. 
Theoretical investigations about this parameter are still lacking so far.
\end{itemize}

Although the above theories provide the relationship between AoA and lift, the influences of the airfoil geometry to the lift curve deserve further investigation.
For example, thin airfoil theory in Eq.\eqref{eq:1} states that $C_{L0}$ is only determined by the airfoil camber. 
Nevertheless, $C_{L0}$ is influenced by other unknown complex factors in engineering practice, and airfoils with the same camber line can exhibit different $C_{L0}$. 
Moreover, the mapping from geometry to some complex characteristics, such as the buffet onset lift coefficient, can hardly be predicted.
Therefore, we aim to establish the relationship between geometry and $C_L$ with neural networks in the present study.

\subsection{Model architecture}
\begin{figure}[htpb]
  \centering
  \includegraphics[width=.3\textwidth]{"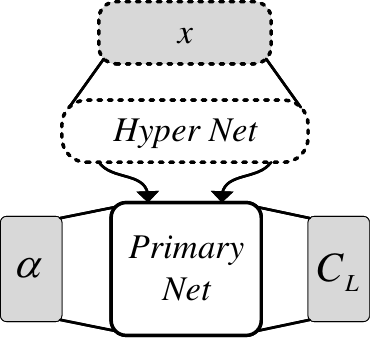"} 
  \caption{Diagram of the proposed model.}
  \label{fig:fig2}
\end{figure}

Consider a supervised learning problem: the airfoil geometry $x\in\mathbb{R}^m$ and AoA $\alpha$ are taken as input to predict the corresponding $C_L$. 
To utilize the theories introduced in Section.~\ref{sec:knowledge}, 
a meta learning model shown in Fig.~\ref{fig:fig2} is proposed, which is composed of a primary network and a hyper network.

The primary network is designed to learn the relationship between $\alpha$ and $C_L$.
Based on the theoretical knowledge, the lift curve can be divided into linear and non-linear portions with a transition point $\alpha_{crit}$.
The linear portion can be represented by the slope $w_{lin}$ and the intercept $b_{lin}$. 
To ensure a proper range for each parameter, $w_{lin}$ and $b_{lin}$ are generated by normalization from the previously discussed $C_{L0}^{\prime}$ and $C_{L0}$, respectively.

For the nonlinear portion, a Multi-Layer Perceptron (MLP) can be used to fit the nonlinear deviation from the linear portion.
Given that the nonlinear portion is expected to be inactive when $\alpha<\alpha_{crit}$, the Rectified Linear Unit (ReLU) activation function $f(x)=max(0,x)$ with $(\alpha-\alpha_{crit})$ as the input is the most obvious choice. 
In this study, we choose the following Softplus function, which is a smooth approximation of the ReLU function, so as to ensure the lift curve is smooth and differentiable, 
\begin{equation}\label{eq:3}
\text{modSoftplus}\left(x\right)=\frac{\log{\left(1+e^{10x}\right)}}{10}
\end{equation}
Then, the primary network is expressed as:
\begin{equation}\label{eq:4}
f(\alpha; \theta)=
\overbrace{w_{lin} \alpha+b_{lin}}^{\text{linear}}+\overbrace{f_{_\text{MLP}}\left(\text{modSoftplus}(\alpha-\alpha_{crit}\right); W_p)}^{\text{non-linear}}
\end{equation}
Where $f_{_\text{MLP}}: \mathbb{R}\mapsto\mathbb{R}$ is a MLP with its weights and biases denoted as $W_p\in\mathbb{R}^{N_{Wp}}$. 
The unknown parameters $\theta$ in the primary network includes $w_{lin}, b_{lin}, \alpha_{crit}$ and $W_p$.

To explicitly encode the airfoil information, a second network (denoted as hyper network) is used to learn the mapping from airfoil geometries to the unknown parameters of the primary network.
The hyper network is given by a MLP network and expressed as follows:
\begin{equation}\label{eq:5}
\theta=g_{_\text{MLP}}(x; W_h)
\end{equation}
Where $g_{_\text{MLP}}: \mathbb{R}^m\mapsto\mathbb{R}^{N_{Wp}+3}$ is a MLP with its weights and biases denoted as $W_h\in\mathbb{R}^{N_{Wh}}$. 

\begin{figure}[htpb]
  \centering
    \subfigure[Primary Net]
    {\includegraphics[width=\textwidth]{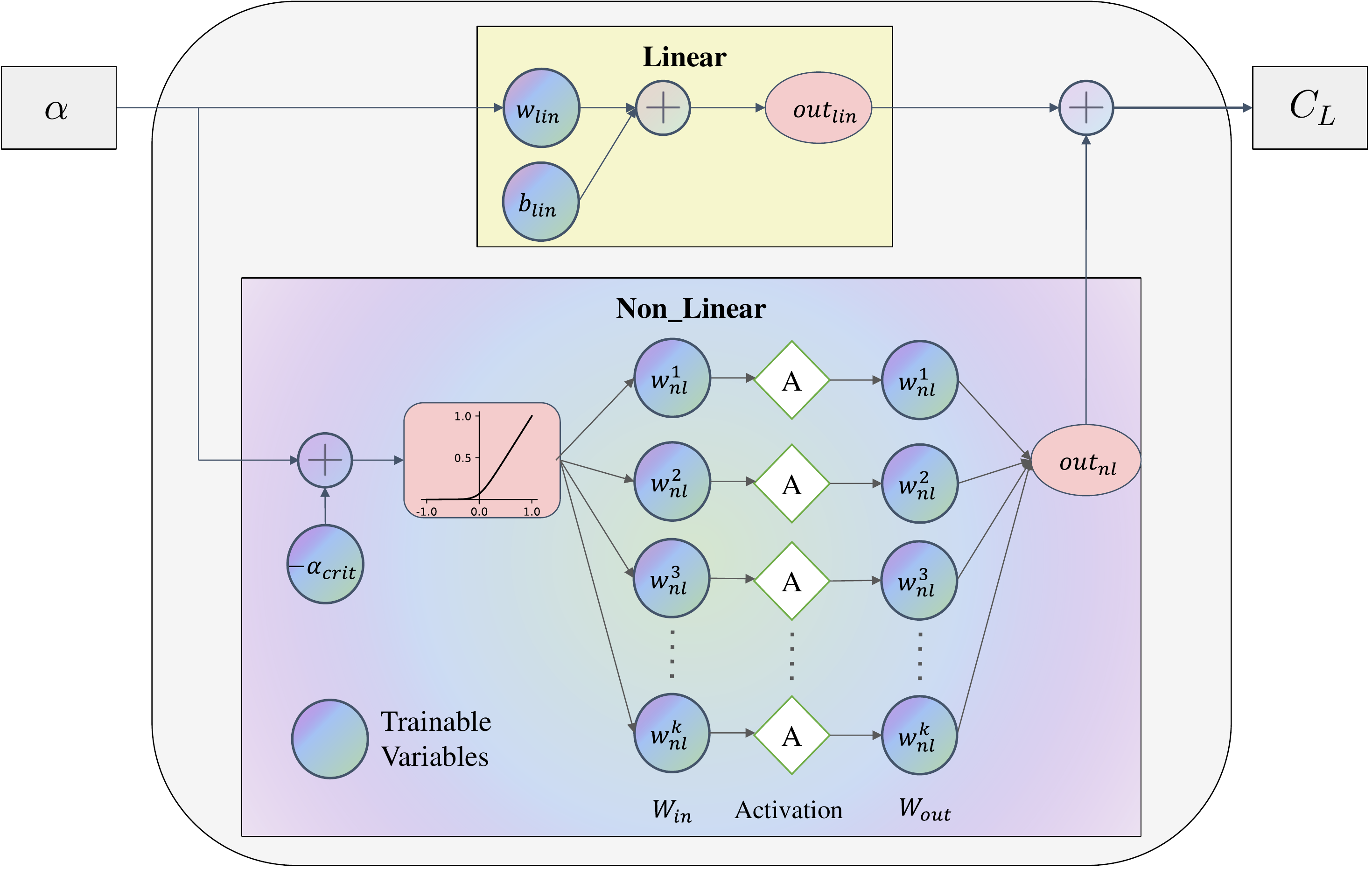}}
    \subfigure[Hyper Net]
    {\includegraphics[width=\textwidth]{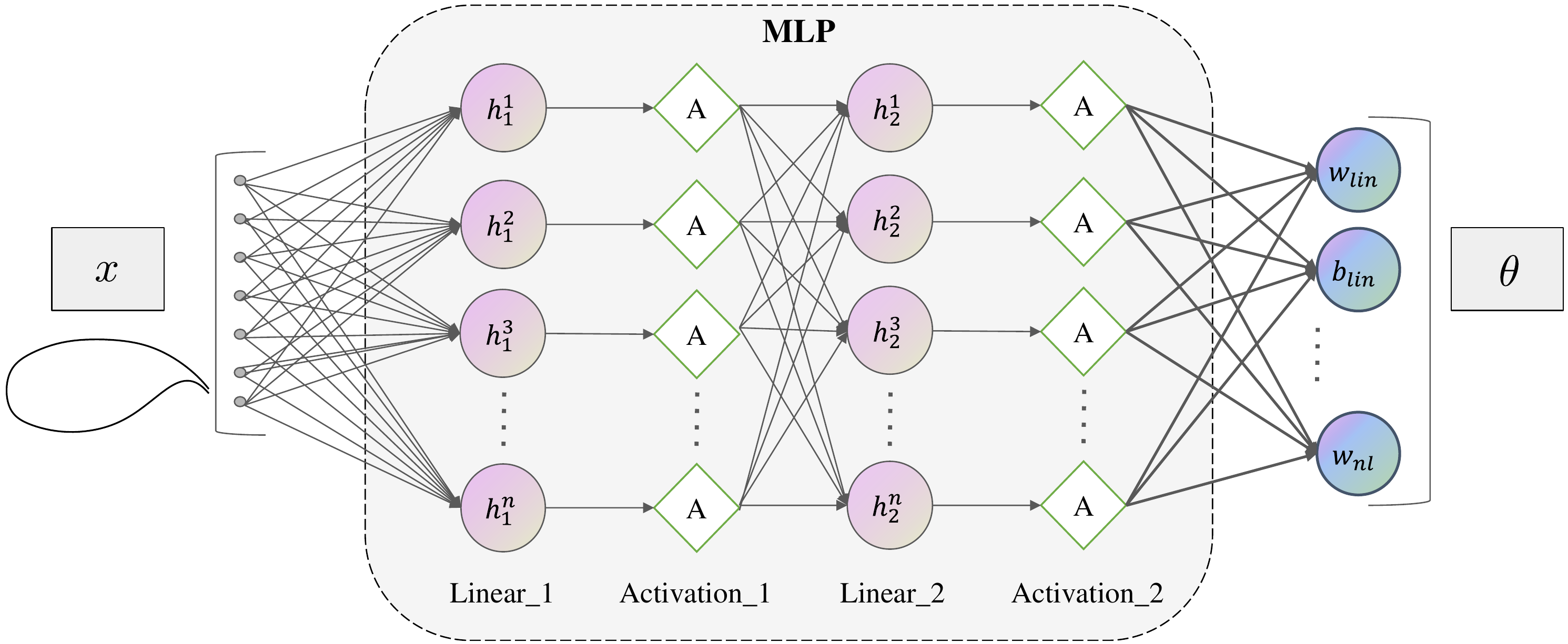}} 
  \caption{Network architecture for the prediction of airfoil aerodynamic performance.}
  \label{fig:fig3}
\end{figure}

The loss calculation of the model follows the meta-learning approach \cite{finn_model-agnostic_2017,ha_hypernetworks_2016}. 
Mean squared error (MSE) is used for this regression tasks. 
The primary net performs the task $\mathcal{T}$ and the task of each airfoil is denoted as $\mathcal{T}_i$. 
Accordingly, the primary net’s parameters are denoted as  $\theta_i$ for each airfoil.
The loss of each task in the primary net takes the form:
\begin{equation}\label{eq:6}
\mathcal{L}_{\mathcal{T}_i}=\sum_{\alpha_{(j)} \sim \mathcal{T}_{(i)}}\left\|f\left(\alpha_{(j)}; \theta_{(i)}\right)-C_{L(j)}\right\|_2^2
\end{equation}
Take the function of the hyper net  into the loss, the network depicted in Fig.~\ref{fig:fig3} is trained by minimizing the following total loss:
\begin{equation}\label{eq:7}
 \mathcal{L}_{total} = \sum \mathcal{L}_{\mathcal{T}_i}=\sum_{\mathcal{T}_{(i)}} \sum_{\alpha_{(j)}}\left\|f\left(\alpha_{(j)}; g\left(x_{(i)}, \phi\right)\right)-C_{L(j)}\right\|_2^2
\end{equation}
In addition, mean absolute error (MAE) and maximum absolute error (MAXE) are also used as evaluation metrics to reflect the prediction accuracy of the model.

\subsection{Dataset}
\begin{figure}[htpb]
	\centering
	\subfigure[Sampled airfoil geometries.]
	{\includegraphics[width=0.5\textwidth]{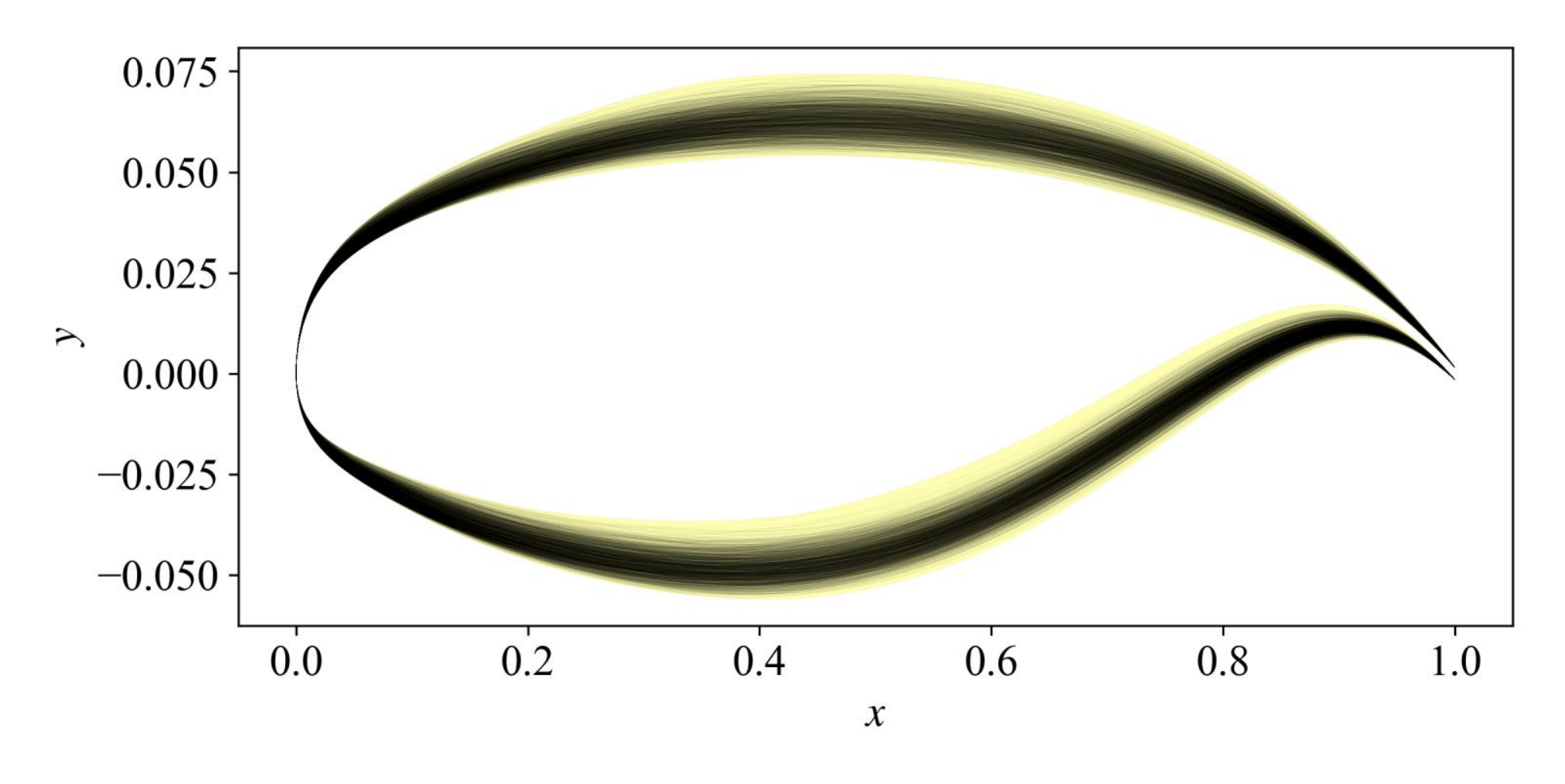}} 
	\subfigure[$C_L$ distribution at different AOAs.] {\includegraphics[width=0.4\textwidth]{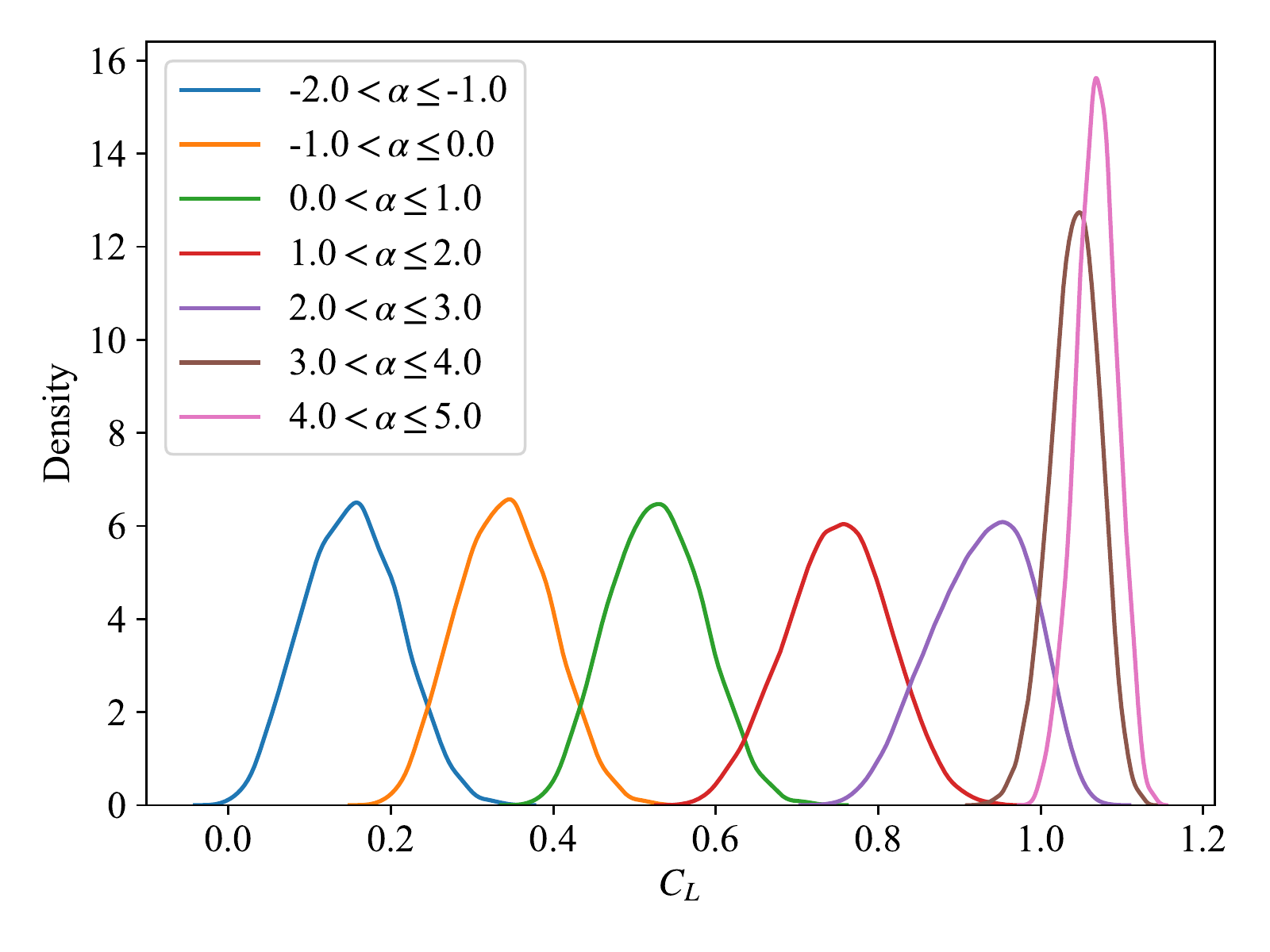}}
	\caption{Dataset details.}
	\label{fig:fig4}
\end{figure}

The dataset includes airfoil geometries and their corresponding $C_L-\alpha$ curves at $Ma=0.73, Re=5.0\times10^6$.
To ensure the availability and diversity of the dataset,  several constraints have been applied during sampling the airfoil geometries, such as the leading edge radius must not be smaller than 0.007 and the maximum thickness of airfoil should be limited in a certain range.
In our study, a total number of 2000 airfoils are sampled with 80\% of the airfoils used as the training set and 20\% as the test set.
To encourage the network to learn the explicit parameters of $C_L-\alpha$ curve accurately, AoA is sampled intensively and each airfoil contains approximately 50 AoAs.
The sampled airfoils and the distribution of $C_L-\alpha$ are depicted in Fig.~\ref{fig:fig4}. 
As can be seen, the dataset involves a wide range of supercritical airfoils and aerodynamic performances.

\begin{figure}[htpb]
	\centering
	\begin{minipage}[t]{0.45\textwidth}
		\centering
	\includegraphics[width=0.8\textwidth]{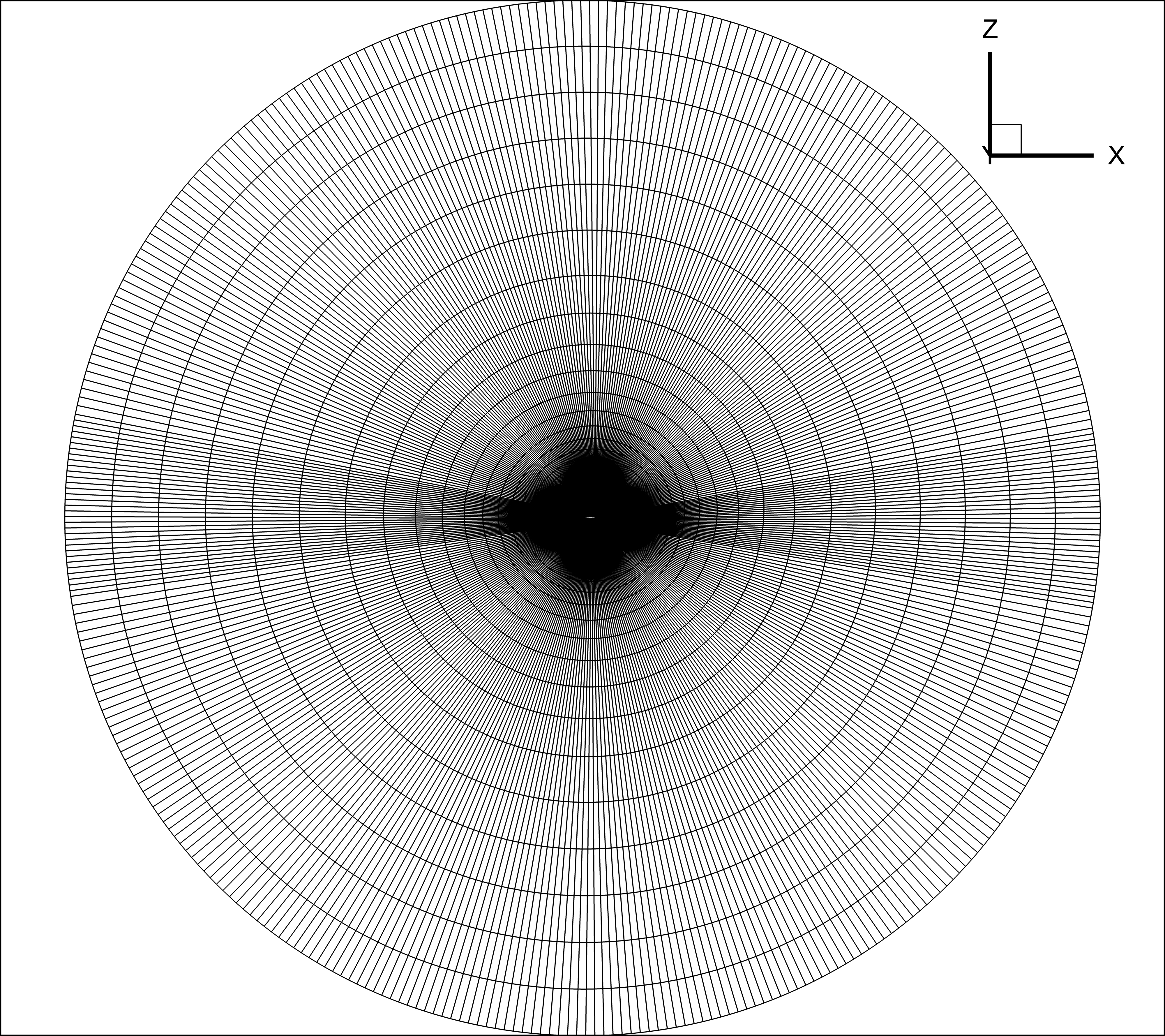}
\end{minipage}
\begin{minipage}[t]{0.45\textwidth}
	\centering
	\includegraphics[width=0.8\textwidth]{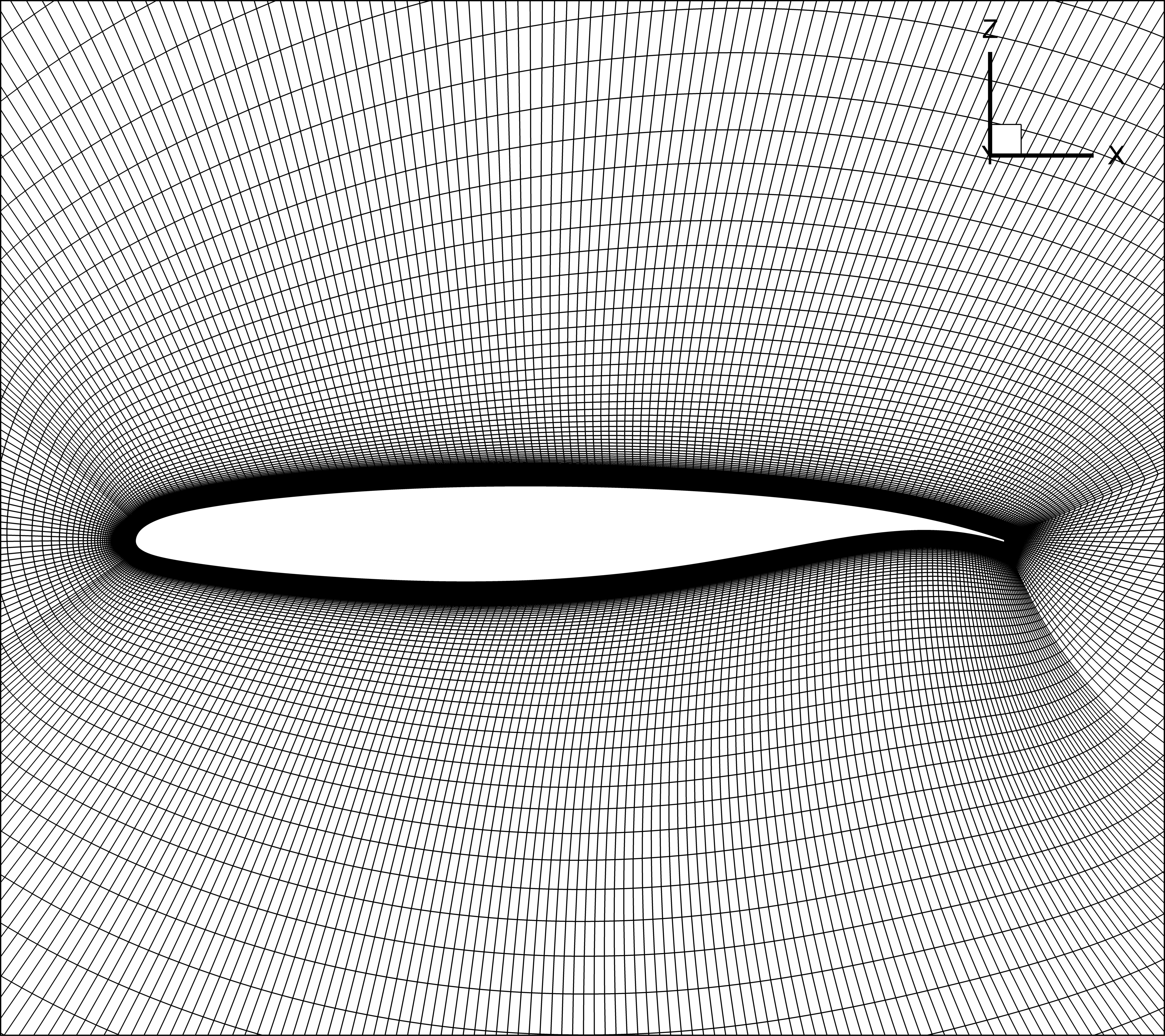}
\end{minipage}
	\caption{The structured O-mesh for numerical simulations.}
	\label{fig:fig5}
\end{figure}

The computational domain is a 2D mesh with a radius of $~30c$, where $c=1.0$ is the chord length of the airfoil.
A structured O-mesh shown in Fig.\ref{fig:fig5} with dimension $385\times193$ is generated in the wrap-around and normal directions, respectively. 
Circumferential grid refinement is performed at the locations where the curvature of the airfoil varies significantly, such as the leading edge and trailing edge. 
By appropriately setting the first layer grid height, the $y^+$ of the airfoil surface is less than 1.

\begin{figure}[htpb]
	\centering
	\subfigure[Comparison of surface pressure with experiment result]
	{\includegraphics[width=0.49\textwidth]{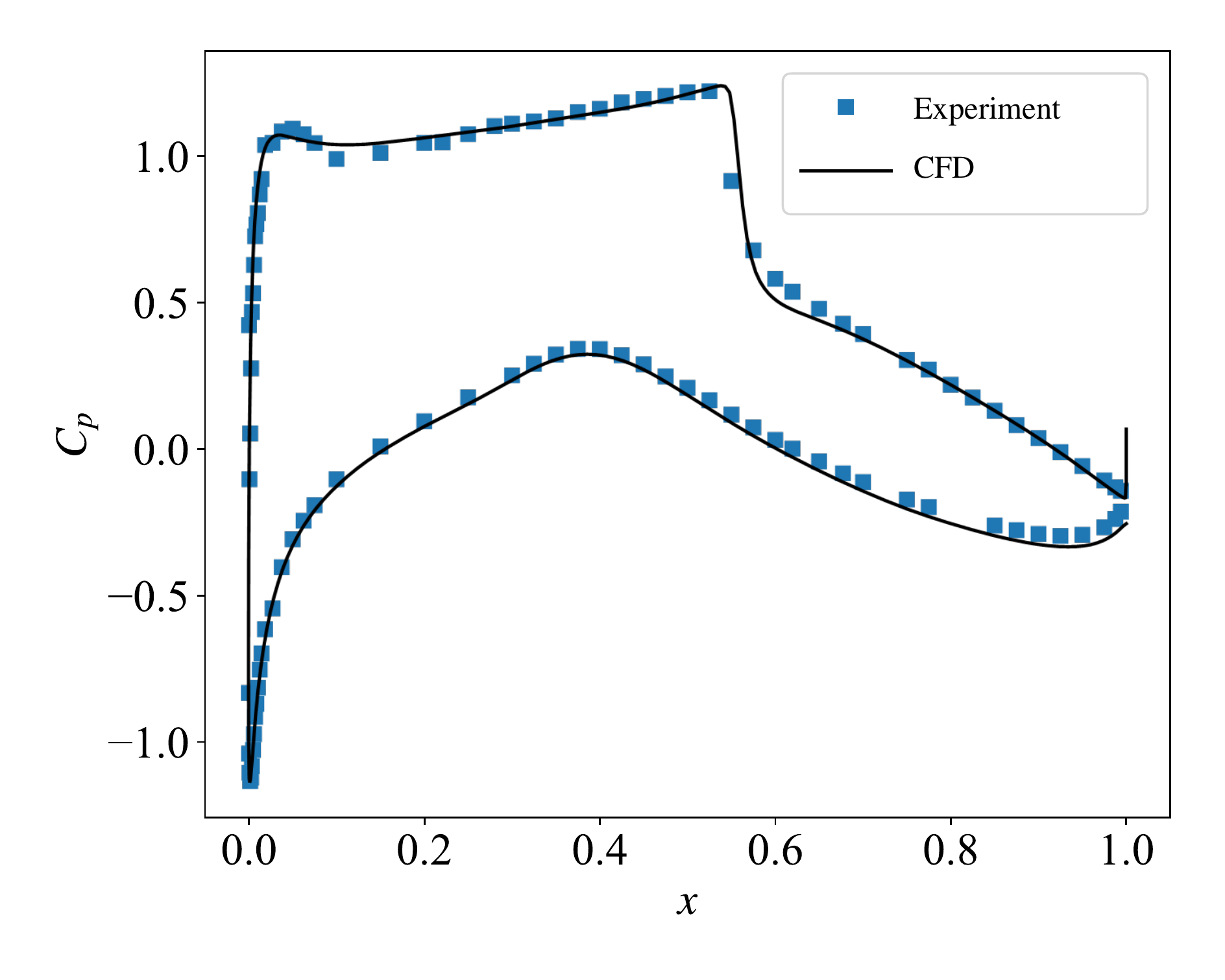}}
	\subfigure[Flow field contour]
	{\includegraphics[width=0.49\textwidth]{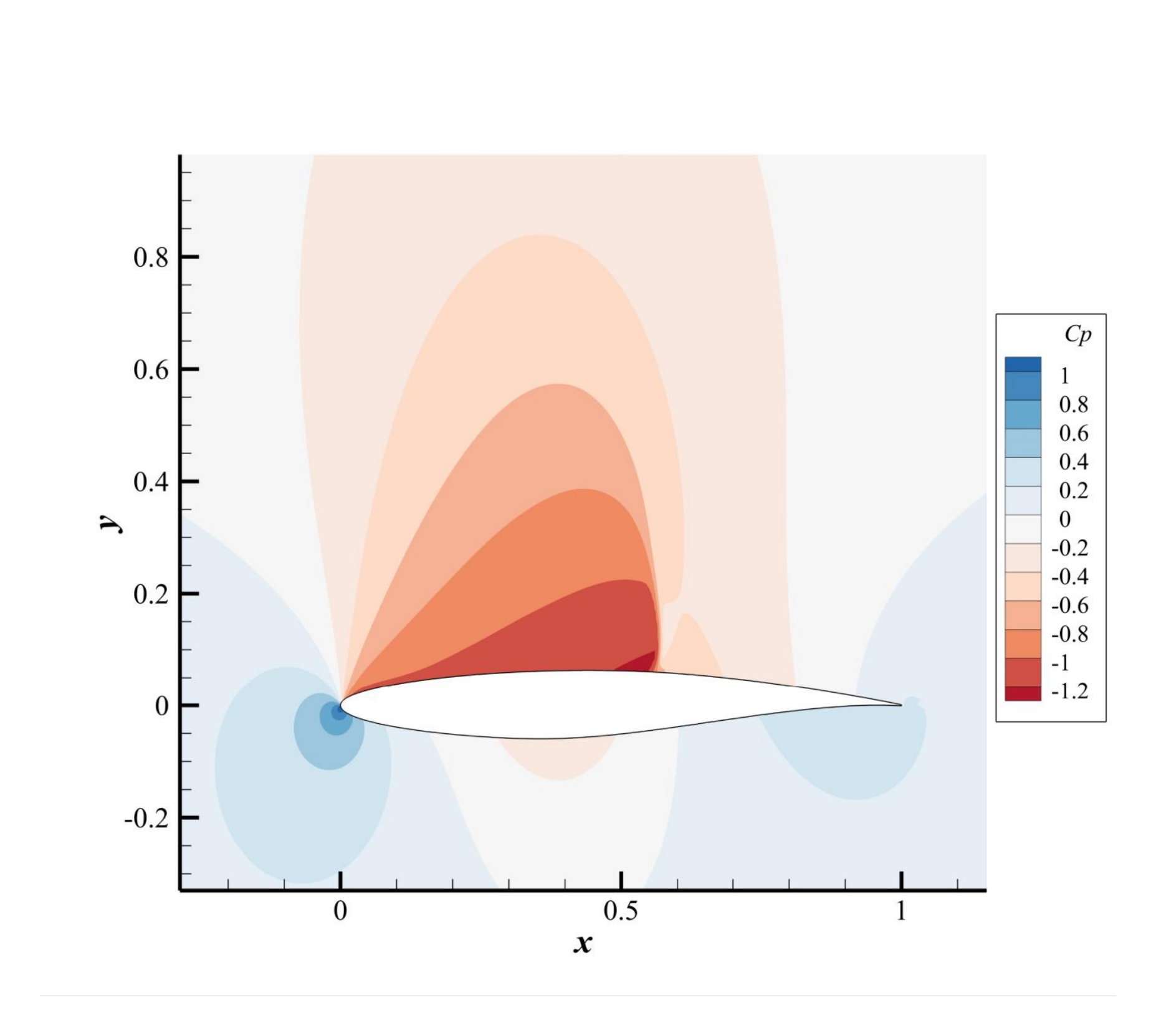}}
	\caption{CFD simulation of the RAE2822 airfoil.}
	\label{fig:fig6}
\end{figure}

The Reynolds Average Navier-Stokes (RANS) simulations make use of the shear stress transport (SST) turbulence model, and solutions are calculated with the open source code CFL3D. 
The monotonic upstream-centered scheme for conservation laws scheme is used to determine state-variable interpolations at the cell interfaces, the Roe scheme is used for spatial discretization and the lower-upper symmetric Gaussian-Seidel method is used for time advancement.
The aerodynamic behaviours of the benchmark RAE2822 airfoil at design flow condition are examined to assess the validity of the numerical scheme. 
The comparison of the pressure distribution between the CFD calculation and the experiment results in FIG.~\ref{fig:fig6}. 
It shows that the CFD calculation can achieve high accuracy, and the flow field contour indicates that the grid settings can capture the main flow structures.

\section{Results and analysis}
The performances of the model are first demonstrated in this section. 
Then, the interpretability of the model and corresponding results are discussed. 
Finally, anomalies with significant errors are further analyzed.

\subsection{Prediction results}\label{subsec:prediction}
In the present study, three models are designed and compared to investigate the performances of the proposed architecture.
(1) \textbf{Base Model:} The primary net takes 1 hidden layer with 10 units followed by the Tanh activation function, and the hyper net takes 2 hidden layers with 128-128 units followed by the Softsign activation function.
(2) \textbf{+Weight Decay:} Weight decay term,  i.e., the sum of the squares of the weights, is added into the loss function of the Base Model, so as to promote sparse weights and simpler relationship between $\alpha$ and $C_L$.
(3) \textbf{-Activation:} Supposing that the relationship between the airfoil geometry and $C_{L0}^{\prime}, C_{L0}, \alpha_{crit}, W_p$ is linear, the activation function of the hyper net in the Base Model is deactivated.

\begin{figure}[htpb]
	\centering
	\subfigure[Training loss]{
		\includegraphics[width=0.75\textwidth]{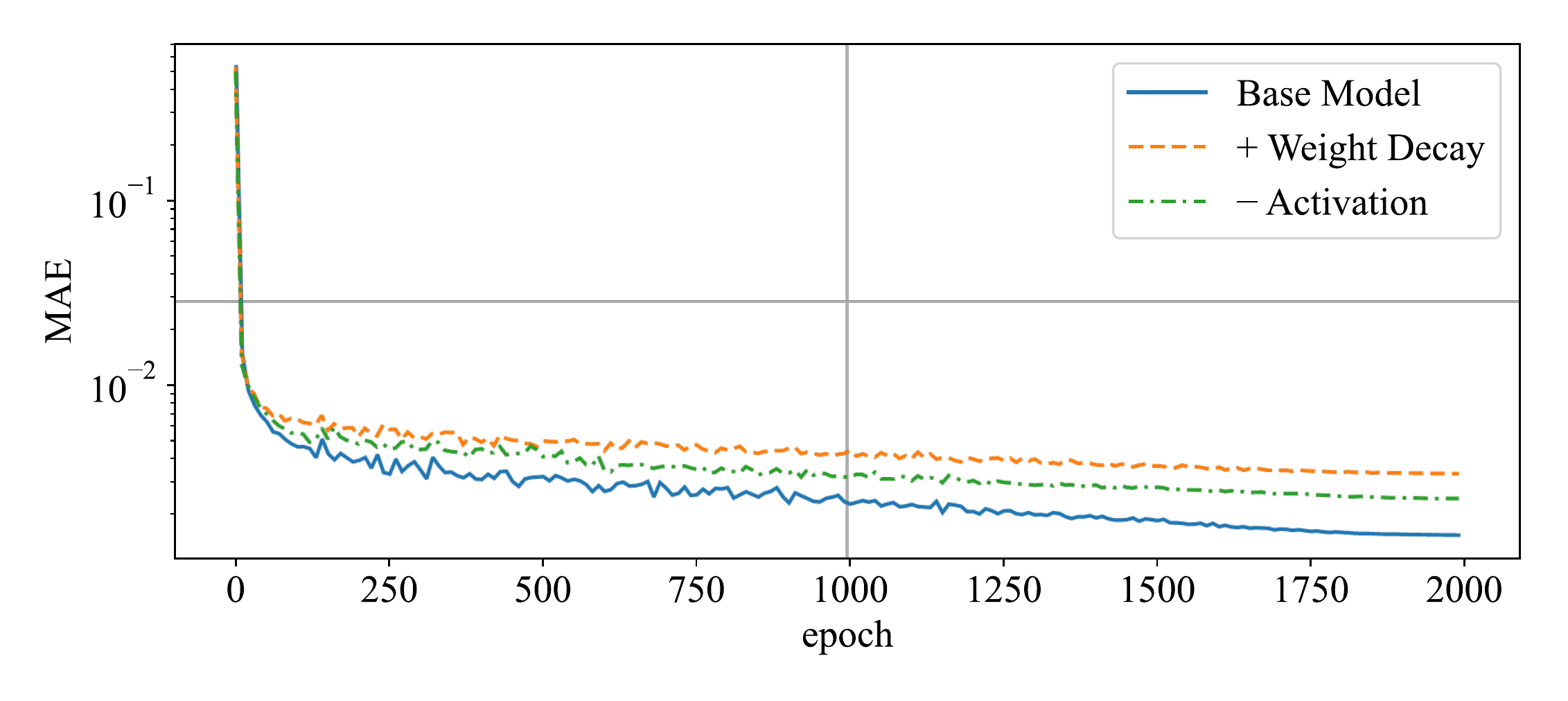}
	}
	\subfigure[Test loss]{
		\includegraphics[width=0.75\textwidth]{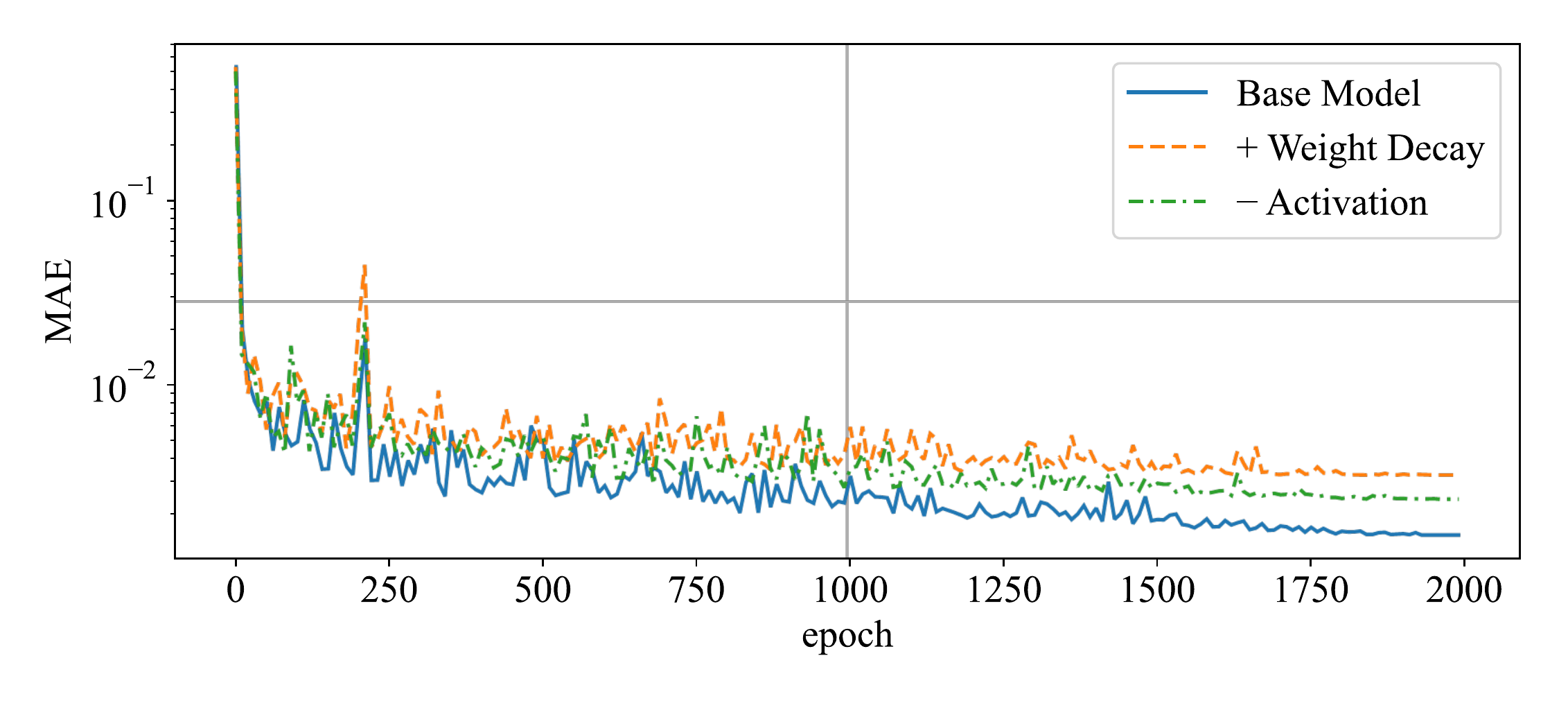}
	}
	\caption{Convergence curves of different configurations.}
	\label{fig:fig7}
\end{figure}

All of the above models are trained with the Adam optimizer.
The learning rate warmup is employed and the initial learning rate is set to $4\times{10}^{-4}$. 
The convergence histories of three models are shown in  Fig.~\ref{fig:fig7}.  
Both the training and testing errors decrease rapidly in the first 250 epochs and then decrease steadily. 
After 2000 epochs, the errors converge at magnitude of ${10}^{-3}$. 
As expected, Base Model achieves the lowest error, followed by -Activation and +Weight Decay. 
The testing loss follows the same trend as the training loss, although the testing curve varies substantially.

To highlight the precision and generalization of the proposed method, we introduce another standard fully connected neural network (denoted as \textbf{ANN}) as comparison, which forms a completely "black box" model with 2 hidden layers with units 258-128-128-1.
The convergence errors of the four models are summarized in Table.~\ref{tab:table 1}. 
MAE reflects the precision of the model,  while MAXE reflects the stability of the model to various samples, indicating how much we can trust the model. 
When comparing the training errors, ANN performs best with the lowest MAE and MAXE.
However, ANN exhibits the largest testing errors, of which the testing MAE is more than twice the training MAE and the testing MAXE is nearly 5 times the training MAXE. 
Similar results can also be found in other studies\cite{yuan2018aerodynamic, du_rapid_2021}, such as 0.36484 for the training loss and 0.06415 for the testing loss.
This indicates that ANN model creates a local mapping confined to the training samples instead of a general and global mapping.

\begin{figure}[htpb]
  \centering
   \includegraphics[width=0.6\textwidth]{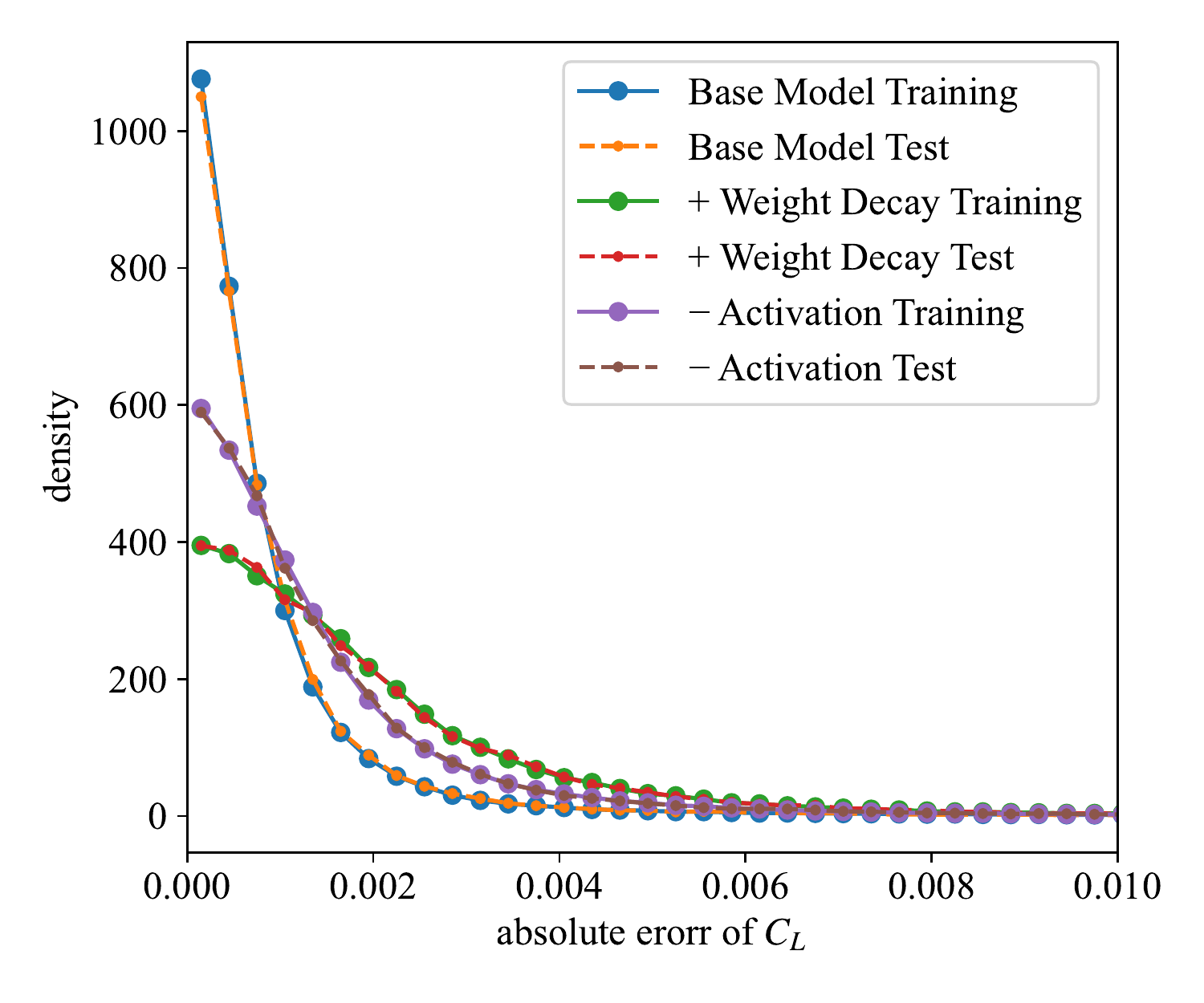}
  \caption{Error distributions of prediction results.}
  \label{fig:fig8}
\end{figure}

In contrast, other three models yields comparable training and testing errors, of which the testing errors are all less than \textbf{ANN}.
For example, the ratios of the training MAE to the testing MAE in Base Model,  +Weight Decay and -Activation are 1.004, 0.984, 0.990, respectively. 
The ratios of the training MAXE to the testing MAXE are 0.885, 0.689, 0.711, respectively. 
Besides, Fig.~\ref{fig:fig8} presents the MAE's probability density distributions of three models, which conform to the normal distribution rule.
The error of training samples has almost exactly the same distribution as testing samples.
As a result of the majority of samples distributed with low MAE, Base Model shows its superior precision over others.
Accordingly, the proposed model exhibits excellent generalization ability and performs equally well on a variety of samples.
The general knowledge embedding in the primary net may be responsible for the significantly improved generalization performance.
Conversely, when presumptions are not met, data may eliminate the discrepancies between knowledge and reality, thus improving the model's accuracy.
Therefore, the dual-driven model achieves comprehensive superiority benefited from both knowledge and data.

\begin{table}[htpb]
\caption{Comparison of different models}
\centering
\begin{tabular}{lcccc}
\hline
     & \multicolumn{2}{c}{Training} & \multicolumn{2}{c}{Test} \\ \cline{2-5} 
     & MAE           & MAXE         & MAE         & MAXE       \\ \hline
	 Base Model & 1.511E-03     & 5.989E-02    & 1.517E-03   & 5.299E-02  \\
	 +Weight Decay & 3.304E-03     & 1.615E-01    & 3.252E-03   & 1.112E-01  \\
	 -Activation   & 2.427E-03     & 1.365E-01    & 2.402E-03   & 9.710E-02  \\ \hline
	 ANN  & 1.467E-03     & 3.282E-02    & 3.457E-03   & 1.611E-01  \\ \hline
\end{tabular}
\label{tab:table 1}
\end{table}

\begin{figure}[htpb]
  \centering
   \includegraphics[page=1,width=0.95\textwidth]{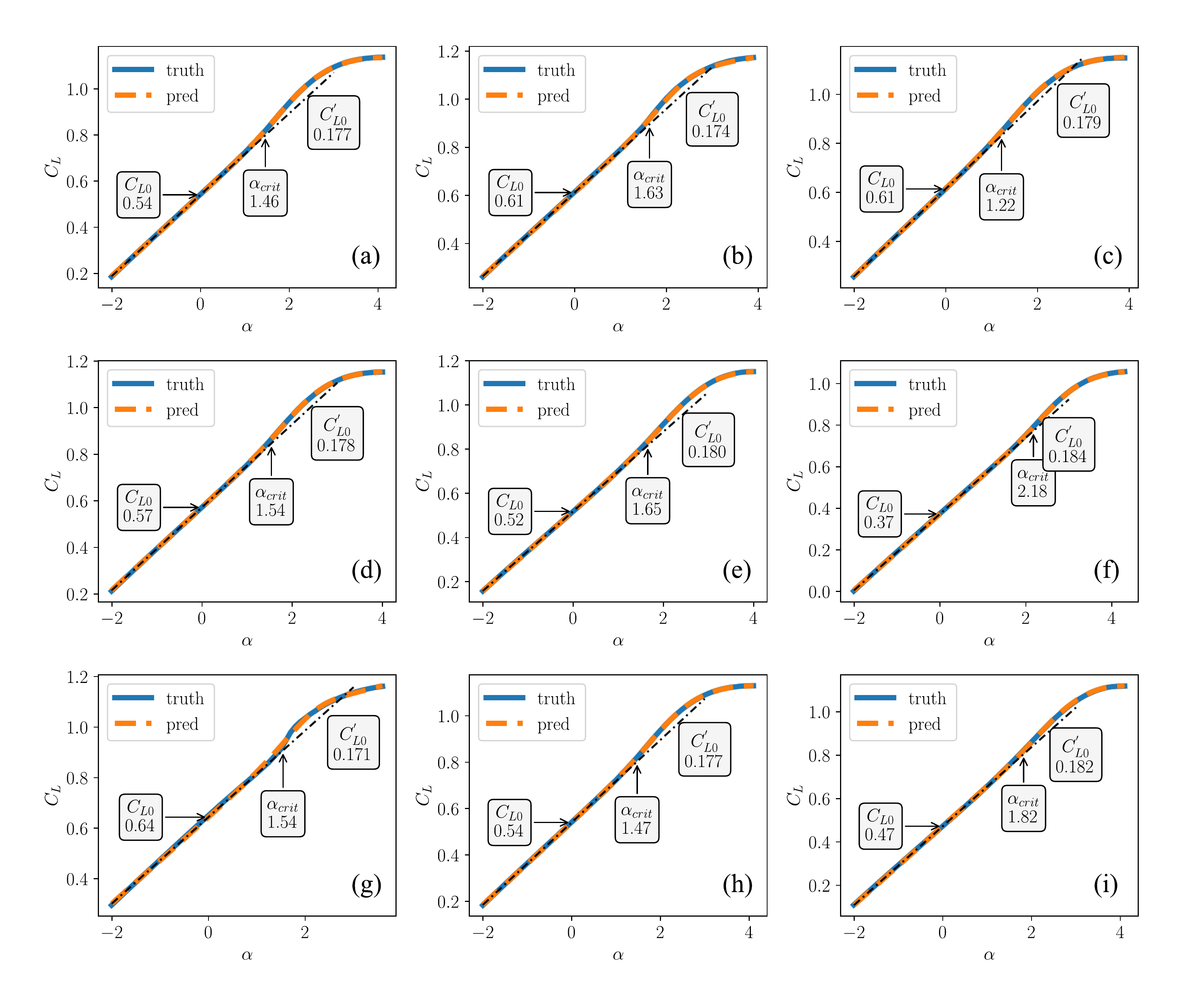}
  \caption{Prediction samples of \textbf{Base Model}.}
  \label{fig:fig9}
\end{figure}

\begin{figure}
  \centering
   \includegraphics[page=2,width=0.95\textwidth]{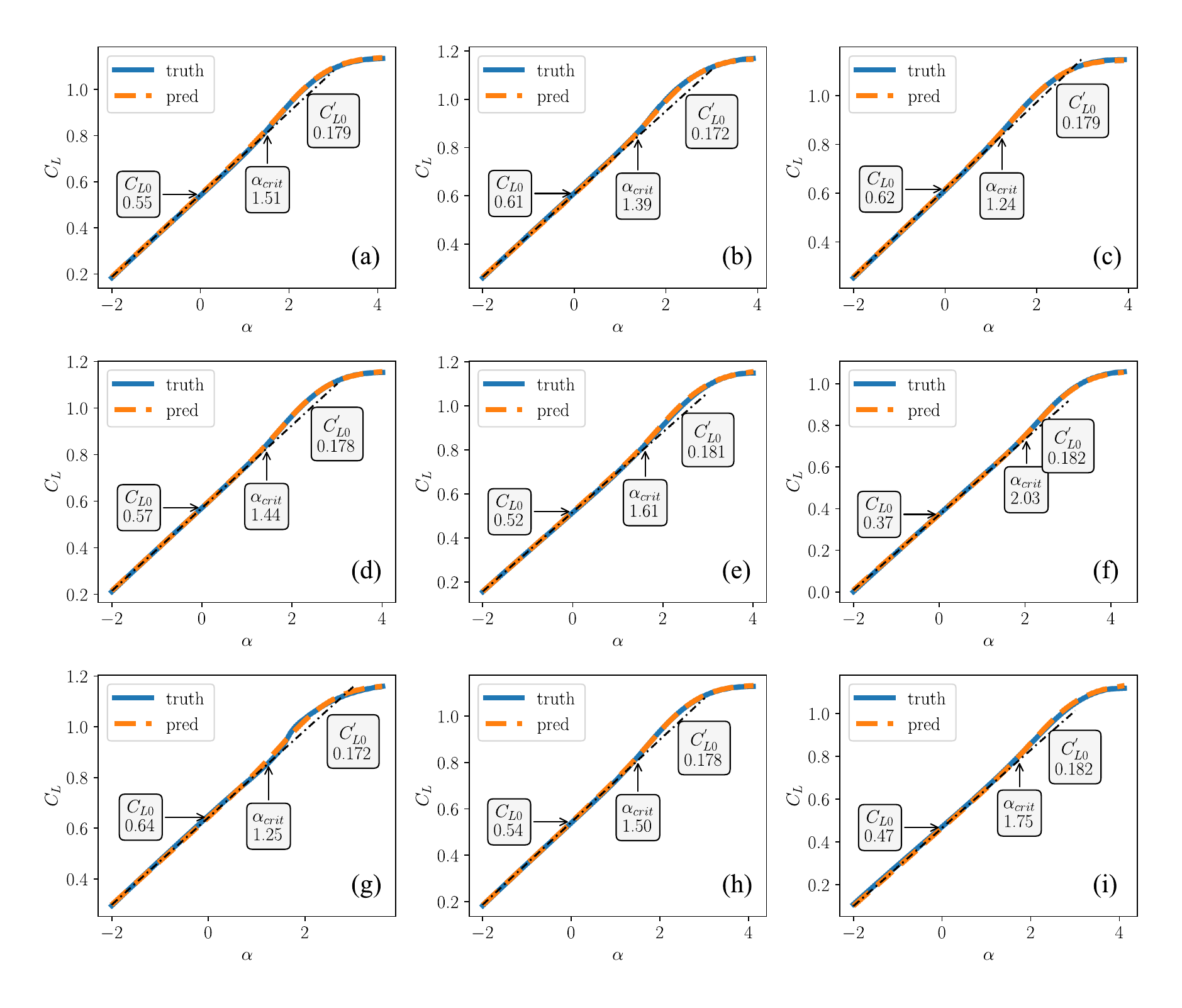}
  \caption{Prediction samples of \textbf{+Weight Decay}.}
  \label{fig:fig10}
\end{figure}

\begin{figure}[htpb]
  \centering
   \includegraphics[page=3,width=0.95\textwidth]{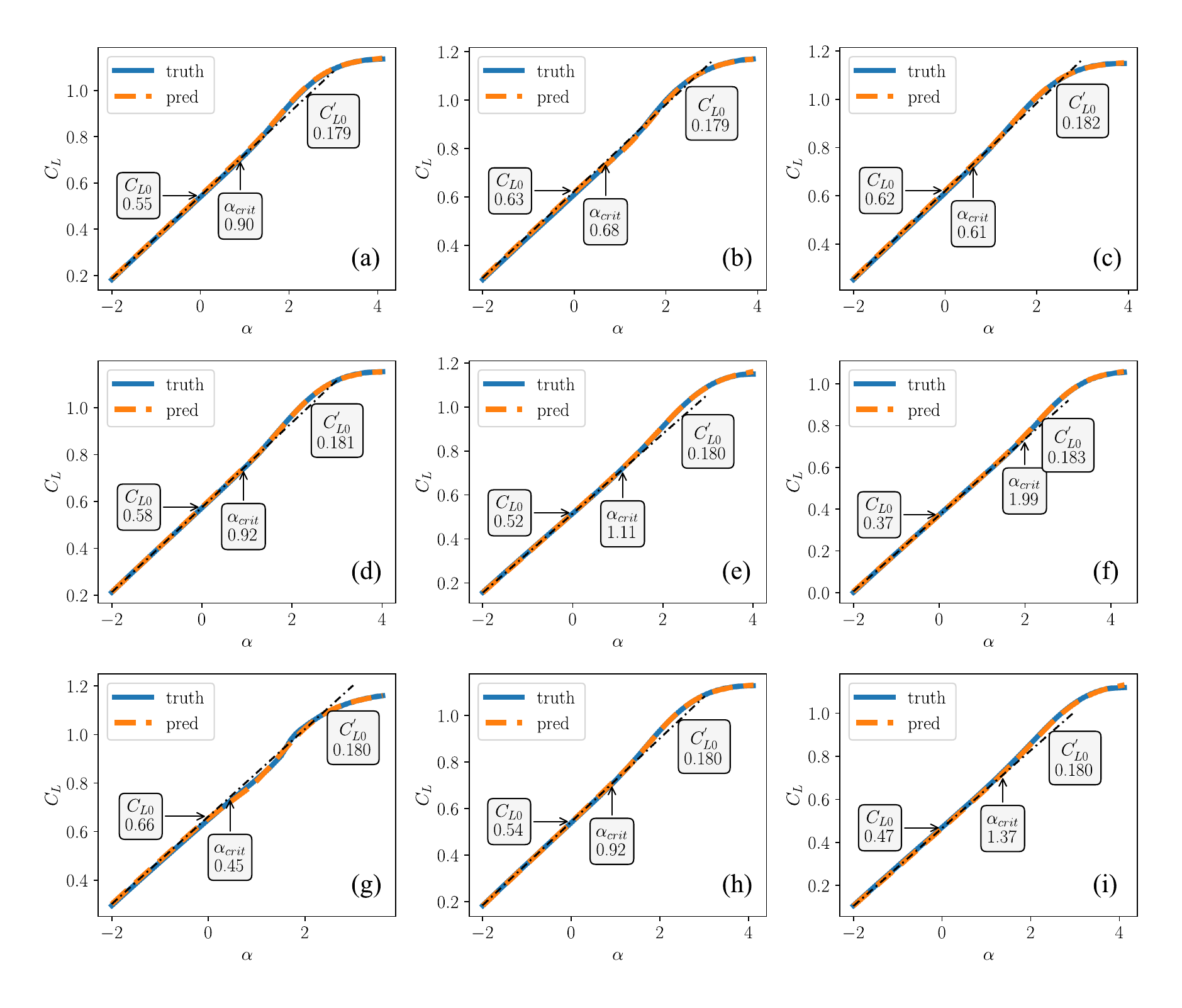}
  \caption{Prediction samples of \textbf{-Activation}.}
  \label{fig:fig11}
\end{figure}

Nine testing samples are randomly selected to demonstrate the performance of the models. 
The predicted lift curves of three models, along with $C_{L0}^\prime$, $ C_{L0}$ and $\alpha_{crit}$ obtained from the hyper net are presented in Fig.~\ref{fig:fig9}-\ref{fig:fig11}.  
The predicted lift curves show a satisfactory match with the actual lift curves for the majority of the samples, where linear and nonlinear portions are all identically fitted.
For \textbf{Base Model}, only sample-g has a discernible error in the nonlinear portion. 
For \textbf{+Weight Decay} model, sample-b/e/g/i have relatively large error in the nonlinear portion. 
For \textbf{-Activation} model, the linear portion of sample-b and sample-g don't fit very well. 
The reason for sample-g's poor prediction is that sample-g locates near the margin of the distribution interval, which can be regarded as an outlier and we will discuss in detail later. 

\begin{figure}[htpb]
  \centering
   \includegraphics[width=0.65\textwidth]{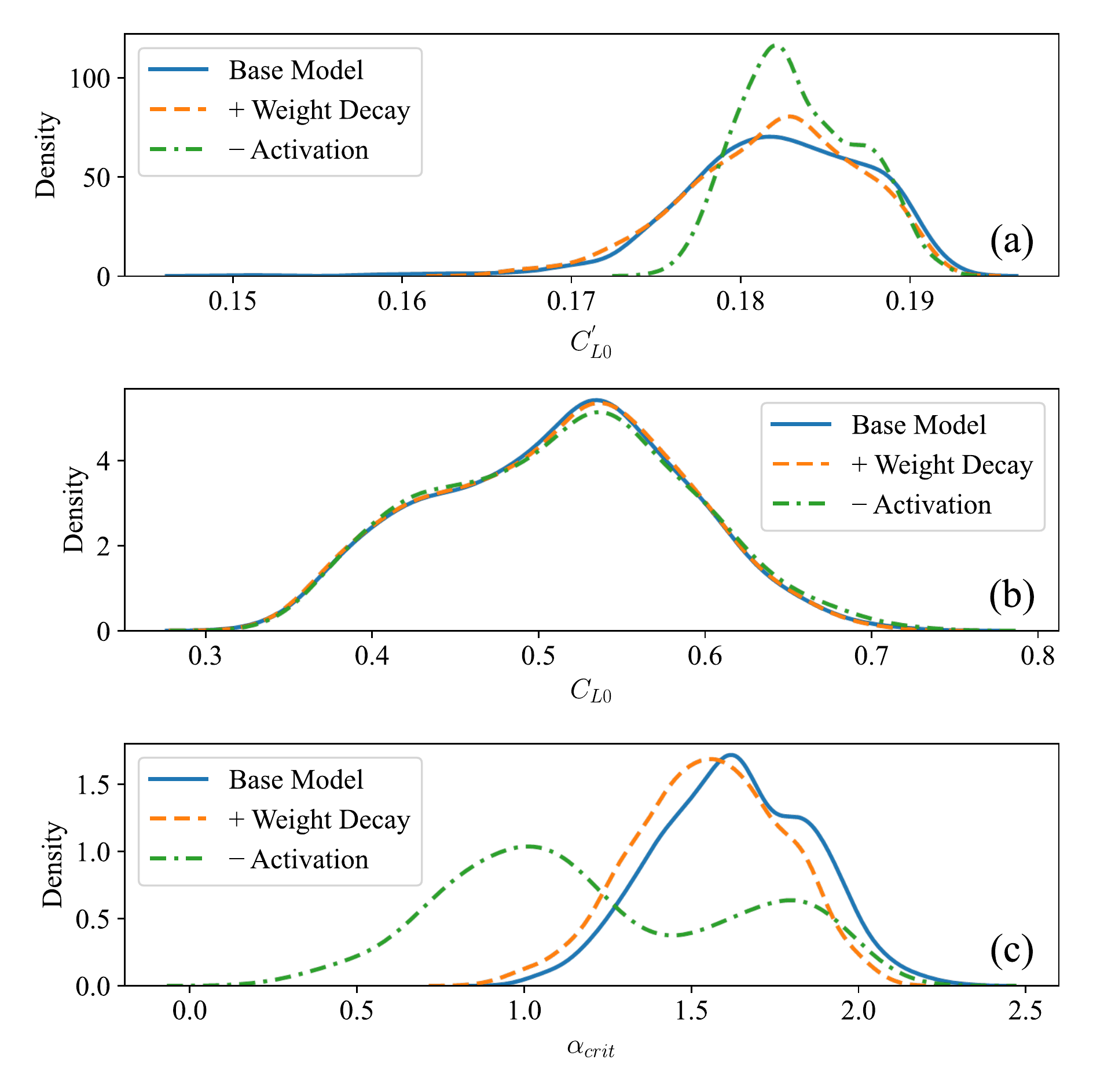}
  \caption{Distribution of key parameters in primary net.}
  \label{fig:fig12}
\end{figure}

The distribution statistics of $C_{L0}^\prime$, $ C_{L0}$ and $\alpha_{crit}$ obtained from the hyper net of three models are presented in Fig.~\ref{fig:fig12}.
For $C_{L0}$, the distributions are quite consistent, indicating that three models achieve almost equivalent prediction. 
The evidence also suggests that the relationship between $C_{L0}$ and geometry is monotonous, which can be expressed through a linear mapping or a nonlinear mapping with sparse weights.
When it comes to $C_{L0}^\prime$ and $\alpha_{crit}$, the distribution of –Activation deviates from the others.
This indicates that the relationship between $C_{L0}^\prime$, $\alpha_{crit}$ and geometry is more complicated and nonlinear, making it difficult to be expressed exactly through a linear mapping.   
To our delight, the distribution of +Weight Decay is relatively similar with Base Model, which is useful for understanding the relationship in a simplified manner.

\subsection{Interpretable analysis}

In this section, we attempt to interpret the models and results, so as to understand the model produce the outcomes and to explore whether knowledge can be discovered through the model.
The primary network is designed based on the theoretical knowledge of the lift curve, in which several trainable parameters, i.e., $C_{L0}^{\prime}, C_{L0}, \alpha_{crit}$, are physically meaningful.
The hyper network represents the effect of airfoil geometry to the physical parameters, and three models have been trained to promote interpretability of the relationship.
Base model is difficult to interpret due to its comprehensive non-linear mappings. 
+Weight Decay model achieves sparse weights and captures primary relationships with subsidiary ones omitted.
-Activation model builds a linear mapping from geometry to $C_{L0}^{\prime}, C_{L0}, \alpha_{crit}$, which can be used to study the overall contribution of each coordinate across all samples.

\begin{figure}[htpb]
  \centering
   \includegraphics[width=0.9\textwidth]{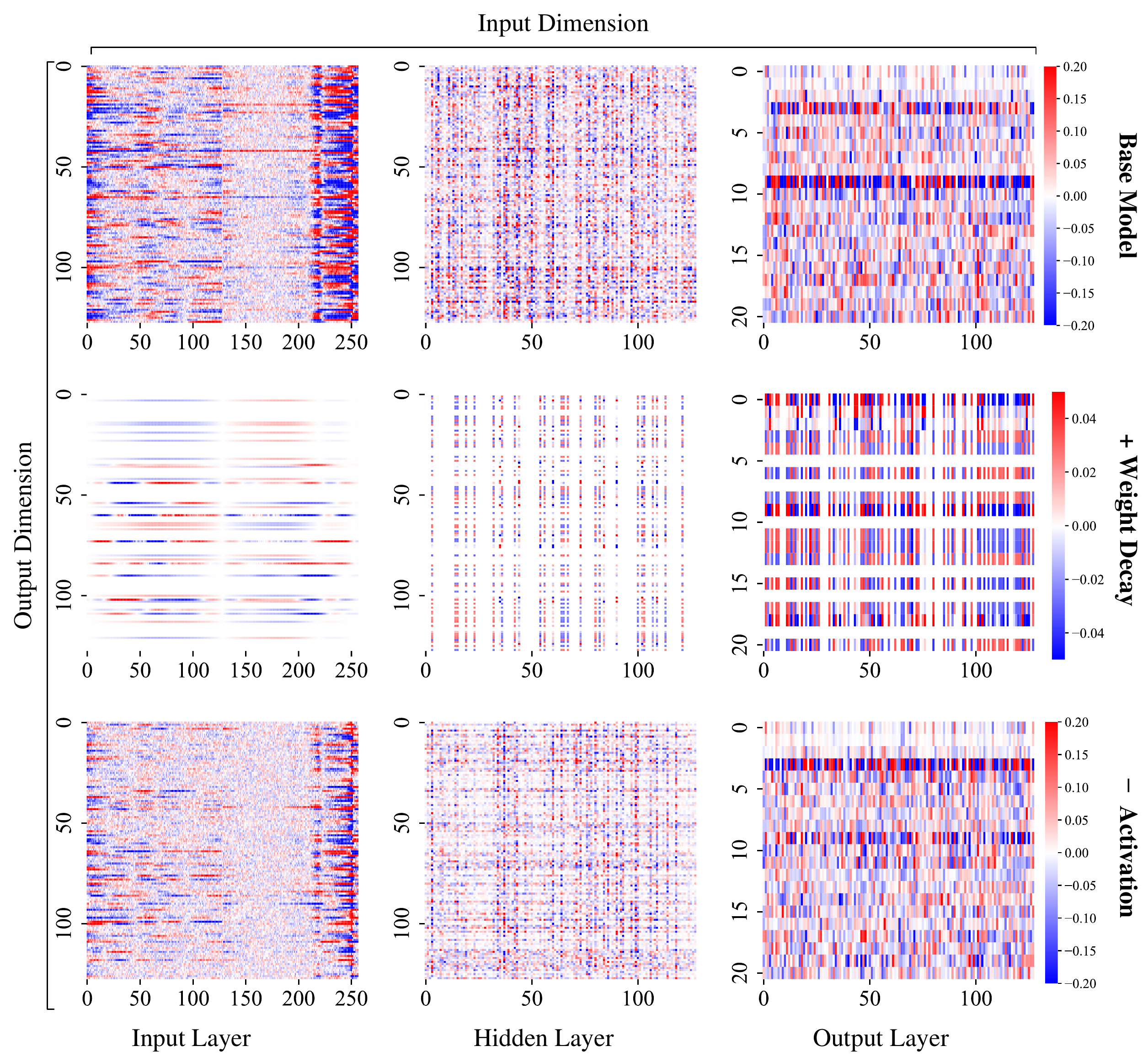}
  \caption{Visualization of the weights of hyper net.}
  \label{fig:fig13}
\end{figure}

Fig.~\ref{fig:fig13} presents the weights of the hyper network in three models. 
The weights of the input layer can reflect the contribution of each part in the geometry to the output.
In Base Model,  it is seen that the upper surface and latter portion of the lower surface have relative larger weights. 
With the addition of the weight decay term, the weights show a significant sparsity and the hyper network is essentially simplified to a smaller scale. 
The -Activation model exhibits the weights highly similar with Base Model.
However, this model has a limited ability to describe local changes in detail due to the lack of activation function. 

\begin{figure}[htpb]
  \centering

    \subfigure[airfoils of variation $C_{L0}$]{
    \includegraphics[width=0.45\textwidth]{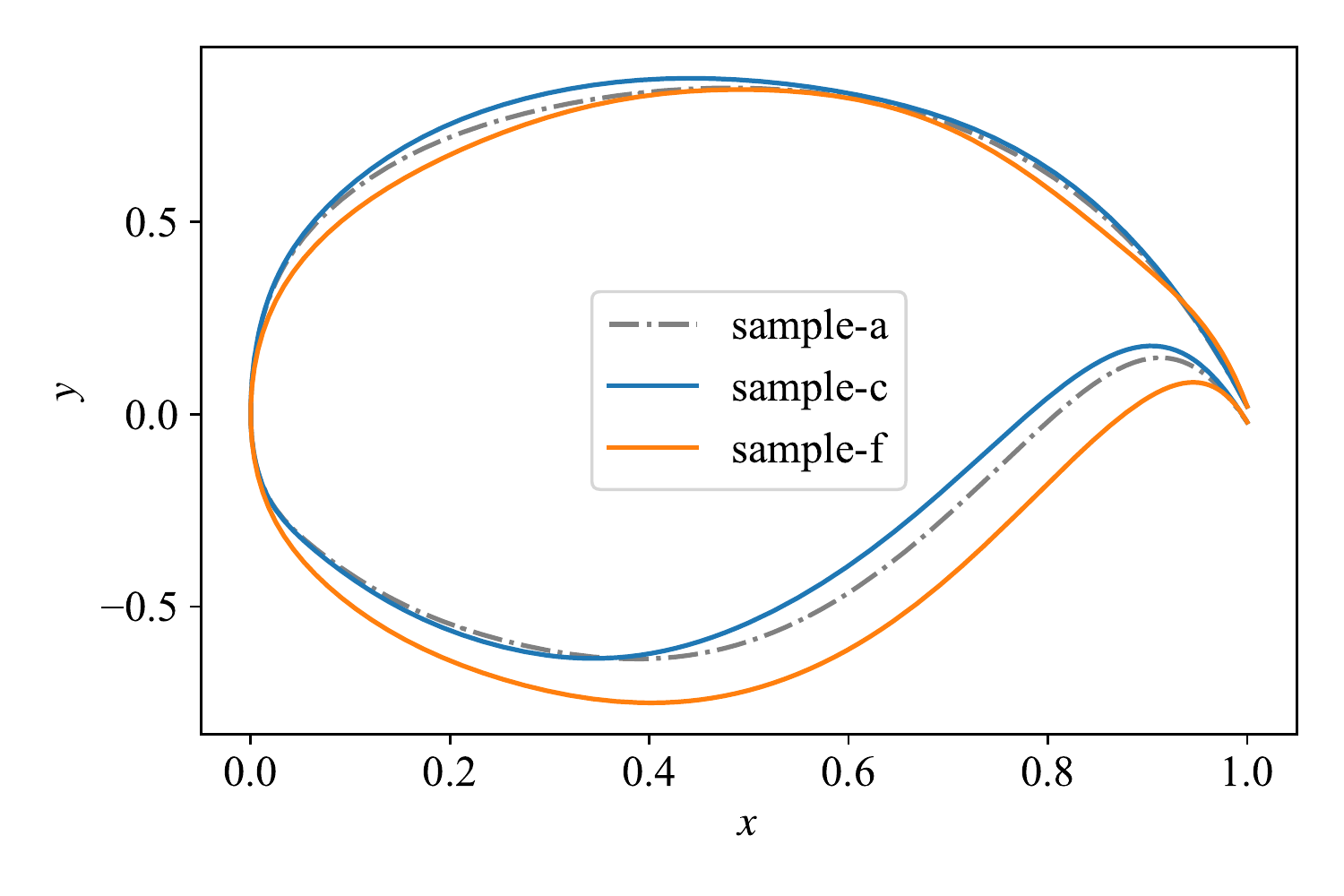}}
    \subfigure[airfoils of variation $C_{L\_crit}$]{
    \includegraphics[width=0.45\textwidth]{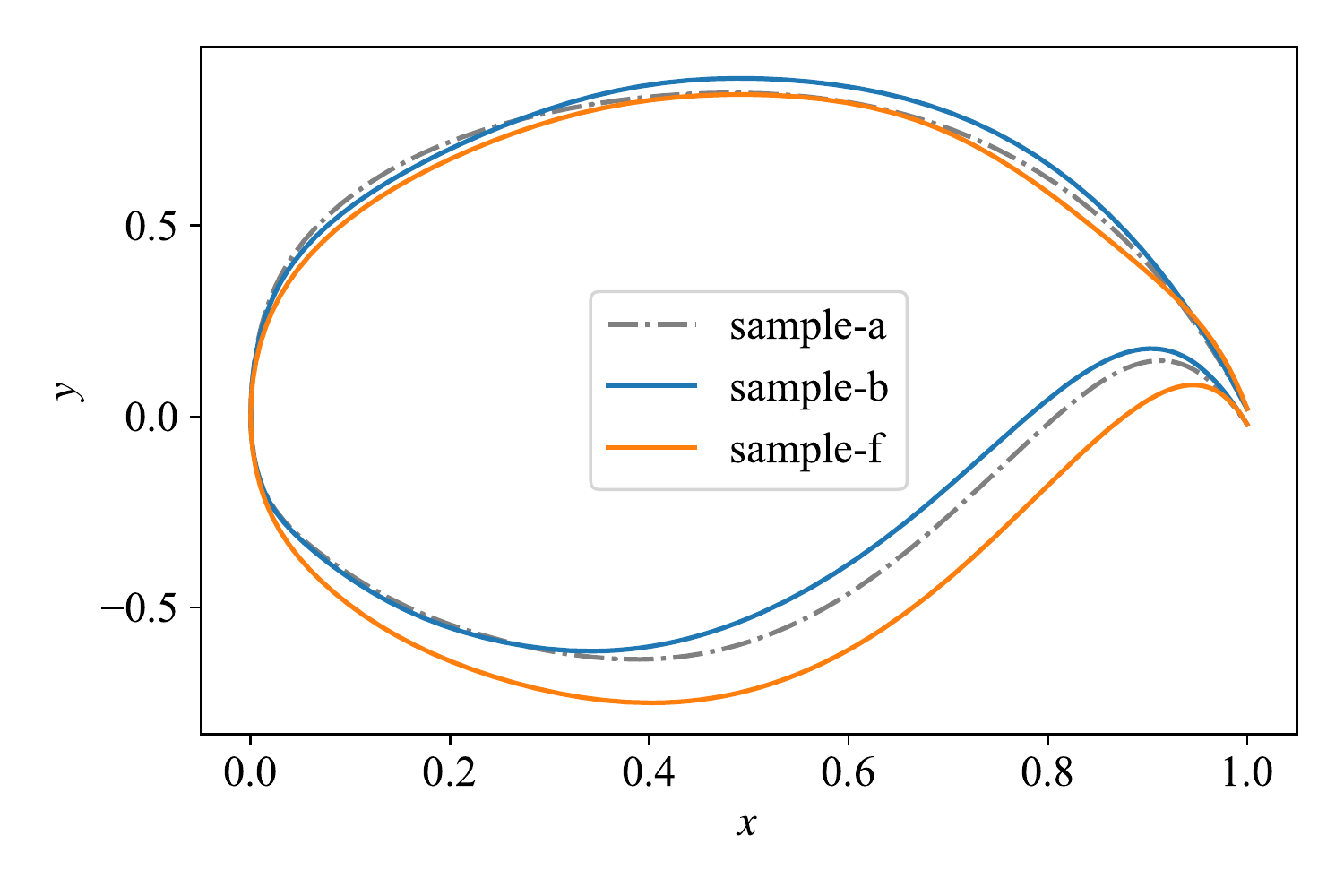}}
    
  \caption{Airfoils selected for interpretability analysis.}
  \label{fig:airfoils_sample}
\end{figure}

To better understand how airfoil geometry affect the physical characteristics, $C_{L0}$ and a new parameter $C_{L\_crit}$, which represents the lift coefficient at $\alpha_{crit}$, are chosen in our study for further analysis.
$C_{L\_crit}$ reflects the maximum lift in the linear portion that can be generated by unseparated flow, which is highly associated with buffet and stall characteristics.
To obtain the contribution of each input to the output, the Integrated Gradients (IG)\cite{Sundararajan_2017_Axiomatic} and Saliency methods are employed in this study. 
IG is an axiomatic model interpretability algorithm to assess the importance score of each input to the output from a given baseline/reference, which satisfies two fundamental axioms: Sensitivity and Implementation Invariance.
Additionally, IG has the completeness that the attributions add up to the difference between the output at the input and the baseline.
Saliency is a baseline approach to obtain the contribution of input through computing the gradients of output with respect to inputs\cite{simonyan_deep_2014}.
Several samples presented in Fig.\ref{fig:airfoils_sample} are selected from the test dataset to conduct IG and Saliency analysis.

\begin{figure}[htpb]
  \centering
  \includegraphics[width=1.0\textwidth]{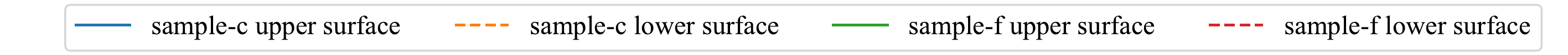}
  \\
    \subfigure[Base Model]{
    \includegraphics[width=0.3\textwidth]{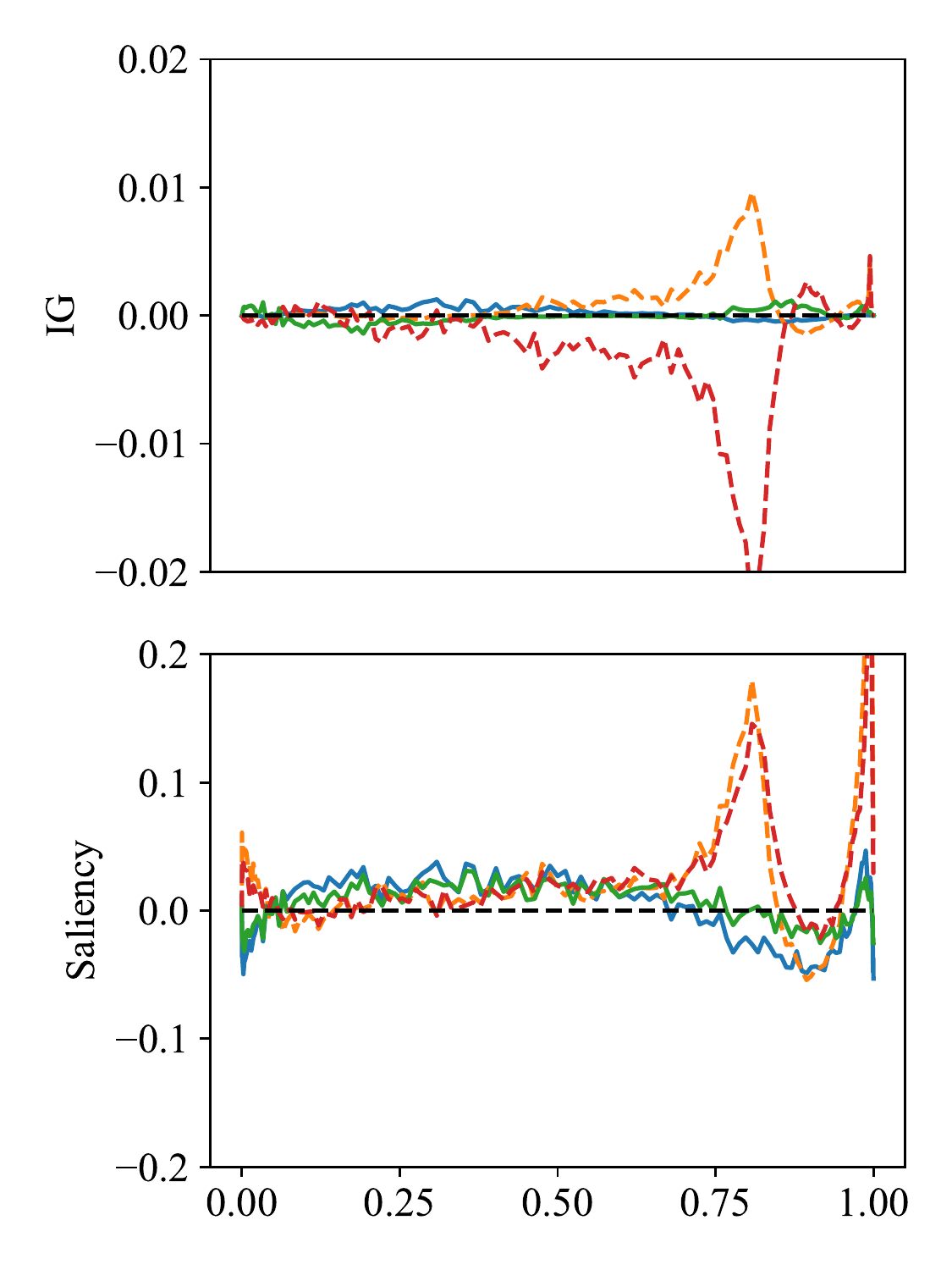}}
    \subfigure[+Weight Decay]{
    \includegraphics[width=0.3\textwidth]{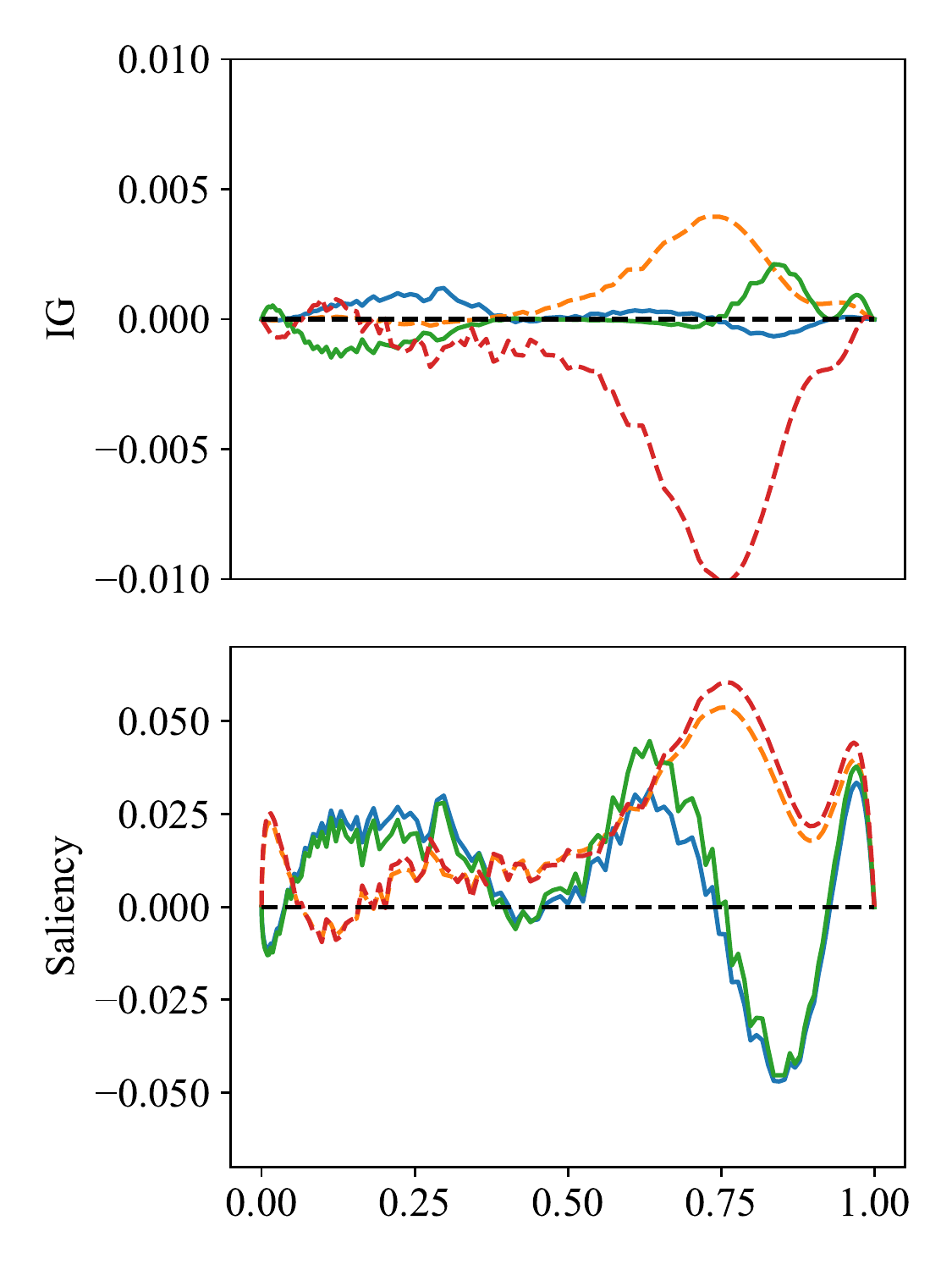}}
    \subfigure[-Activation]{
    \includegraphics[width=0.3\textwidth]{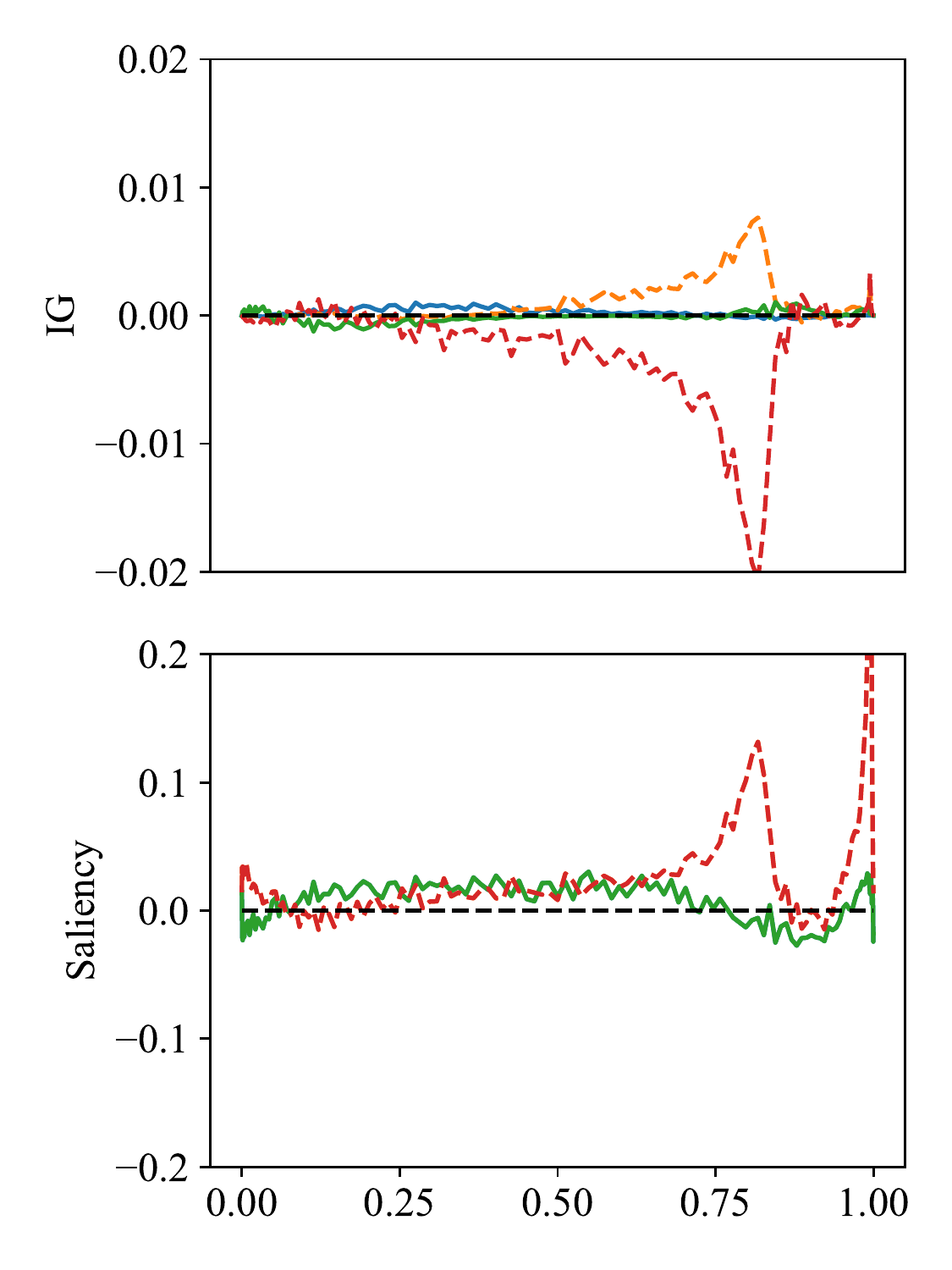}}
  \caption{interpretability analysis of $C_{L0}$.}
  \label{fig:CL0}
\end{figure}

First, we explore the relationship between $C_{L0}$ and geometry.
$C_{L0}$ of sample-a/c/f in Fig.\ref{fig:airfoils_sample}(a) obtained from Base Model are 0.54, 0.61, and 0.37, respectively. 
Taking sample-a as the reference, sample-c exhibits a higher $C_{L0}$ and sample-f exhibits a lower $C_{L0}$.
The input attributions to $C_{L0}$ obtained from three models are illustrated in Fig.\ref{fig:CL0}.
As expected, both IG and Saliency curves obtained from Base Model and -Activation are highly consistent.
+Weight Decay achieves much smoother curve with less oscillation, which is the optimal model to interpret.
As can be seen from the Saliency curves, the latter portion of the lower surface has the greatest positive influence on $C_{L0}$, which corresponds to the evident changes in the IG curves.
Positive attribution of upper surface falls within the range of $0.1c$ to $0.7c$, whereas negative attribution falls within the range of $0.7c$ to $0.9c$.
This can also be observed in IG curves, even more significant impact appears on the lower surface.
For the leading edge portion of the airfoil ($<5\%c$), the upper and lower surface yield opposite attribution in the Saliency curves, which indicates that a smaller leading edge radius is more favorable for $C_{L0}$. 
However, the contribution of the leading edge portion can be ignored, as seen in the IG curves, 
given that minor differences in the absolute value of Saliency and thickness.
The above findings imply that all variations of the upper and lower surfaces are closely related to the airfoil's camber line, particularly in light of the crucial fact that the lower surface has a greater impact on the curvature of the supercritical airfoil.

\begin{figure}[htpb]
  \centering
  \includegraphics[width=1.0\textwidth]{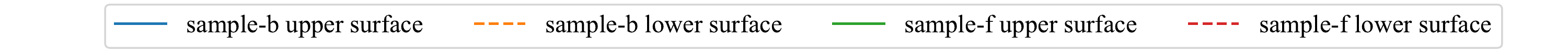}
  \\
    \subfigure[Base Model]{
    \includegraphics[width=0.3\textwidth]{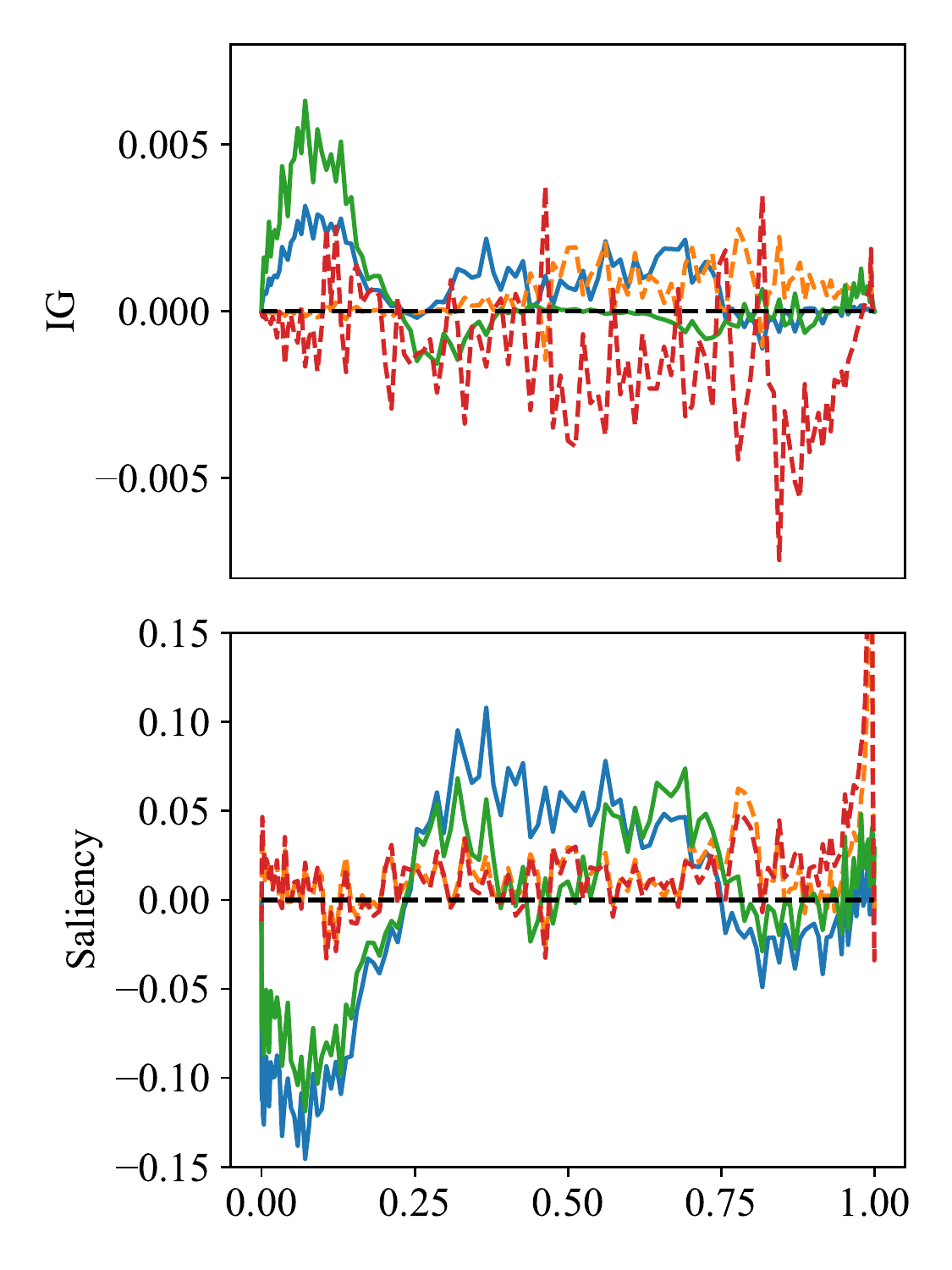}}
    \subfigure[+Weight Decay]{
    \includegraphics[width=0.3\textwidth]{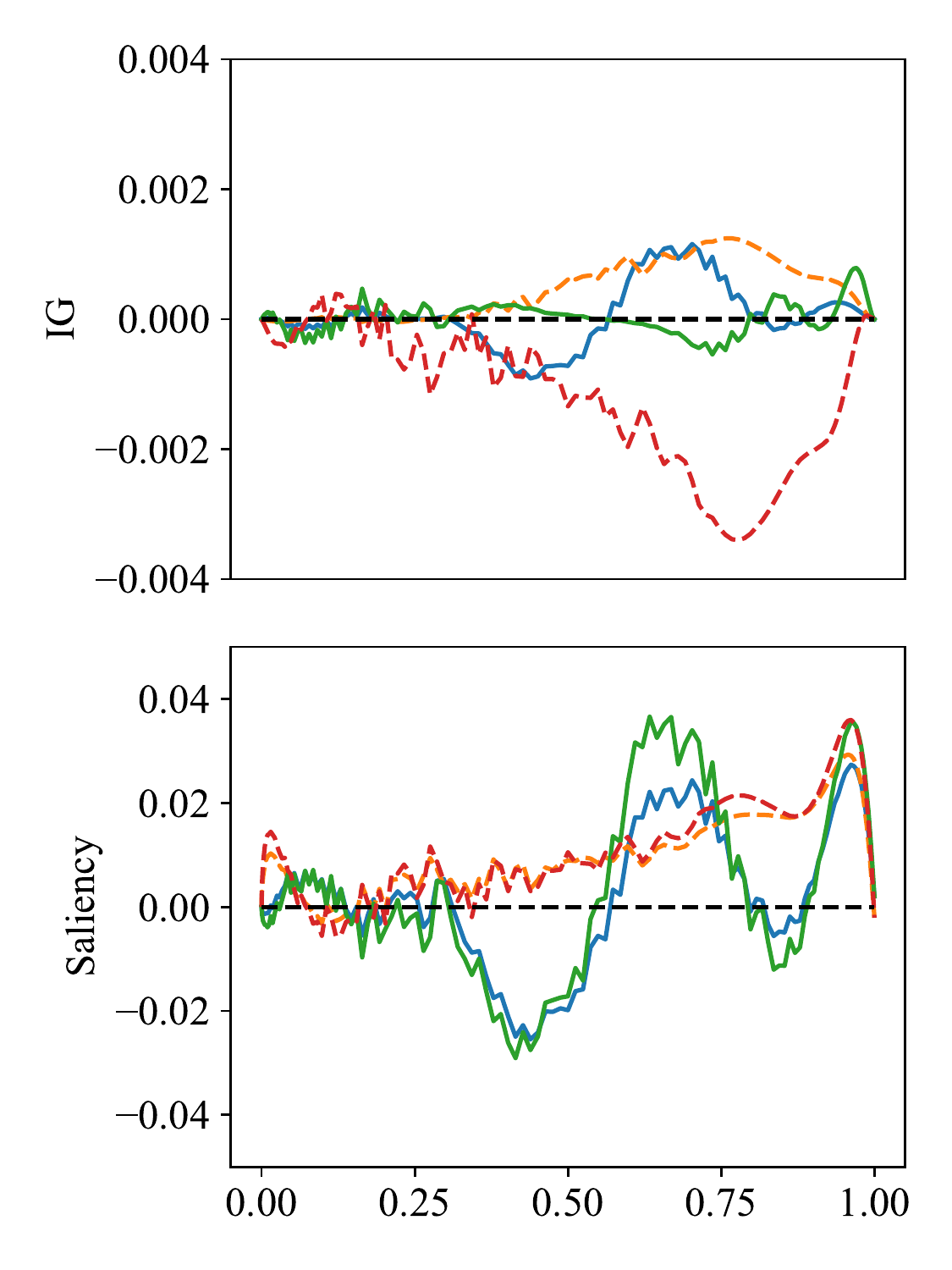}}
    \subfigure[-Activation]{
    \includegraphics[width=0.3\textwidth]{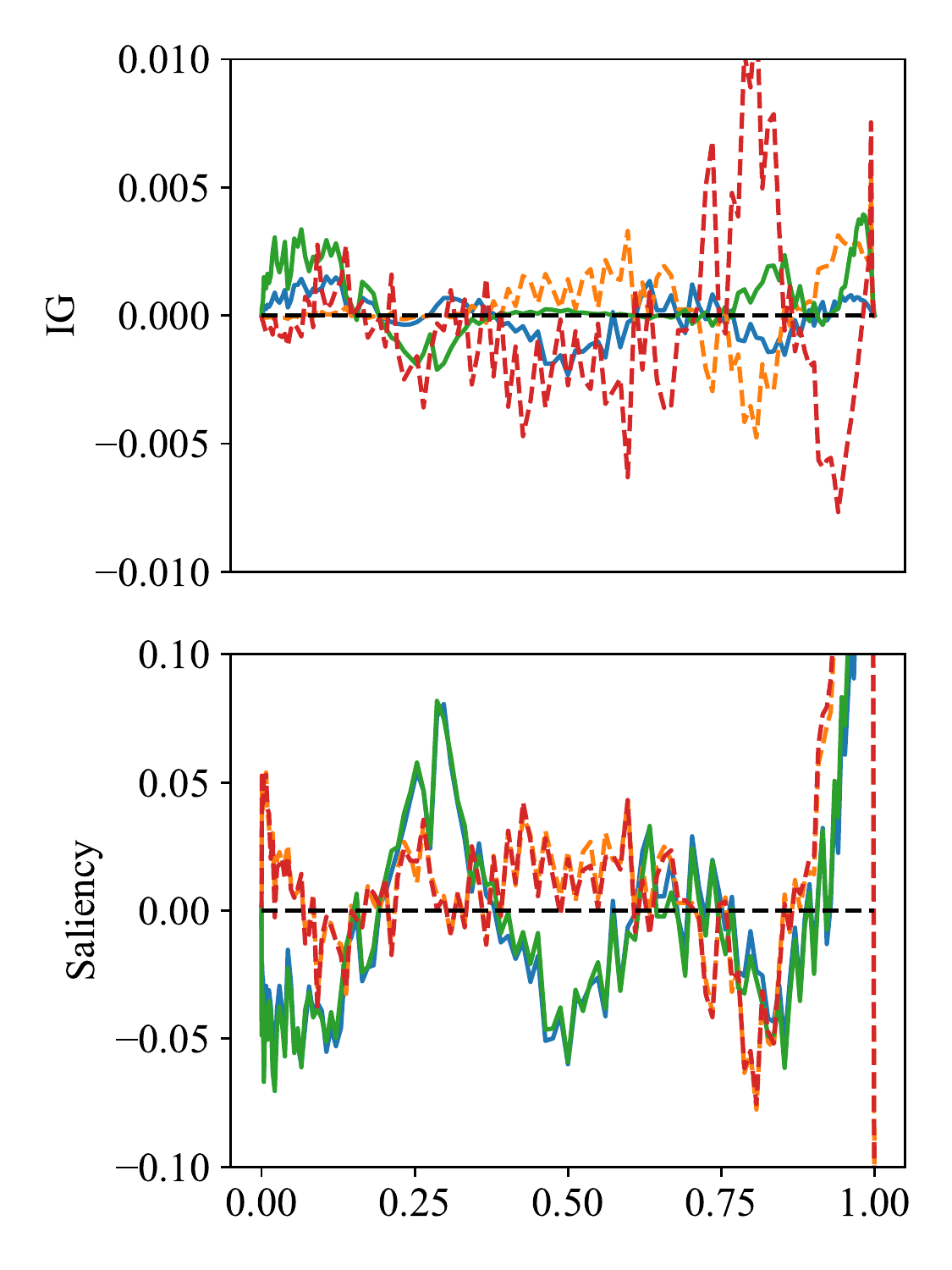}}
  \caption{interpretability analysis of $C_{L\_crit}$.}
  \label{fig:CLcrit}
\end{figure}

Then, we explore the relationship between $C_{L\_crit}$ and geometry.
$C_{L\_crit}$ of sample-a/b/f obtained from Base Model are 0.7984, 0.8956, 0.7724, respectively.
Taking sample-a as the reference, sample-b exhibits a higher $C_{L\_crit}$ and sample-f exhibits a lower $C_{L\_crit}$.
The input attributions to $C_{L\_crit}$ obtained from three models are illustrated in Fig.\ref{fig:CLcrit}.
Unlike $C_{L0}$, the IG and Saliency curves of $C_{L\_crit}$ obtained from three models vary with each other.
As expected, Base model learns the most comprehensive relationship allowing for local adaptation to different samples.
-Activation model exhibits an identical Saliency curve across various samples, which can only be stated in a sophisticated global manner.
+Weight Decay model preserve the primary impacts through eliminating minor ones such as the leading edge section of the airfoil upper surface, which is the optimal model to interpret.
As can be seen from the Saliency curves, the lower surface of the airfoil has a positive impact on  $C_{L\_crit}$.
When it comes to the IG curves, the lower surface of samples-b and samples-f yield opposite attributions given that they change in opposite directions.
For the upper surface near the leading edge portion ($<25\%c$), negative attribution is found in the Saliency curves obtained from Base model, which is caused by the shifts down of sample-b and sample-f in this region.
For sample-f, the contribution of the lower surface is more evident than that of the upper surface,  resulting in a lower $C_{L\_crit}$ than sample-a.
Based on the above analysis, the proposed model can exhibit local interpretation for various geometries, which can provide guidance for local optimization problems.

\subsection{Exceptions}

\begin{figure}[htpb]
  \centering
    \subfigure[$\alpha=-2.0^{\circ}$]{
    \includegraphics[width=0.3\textwidth]{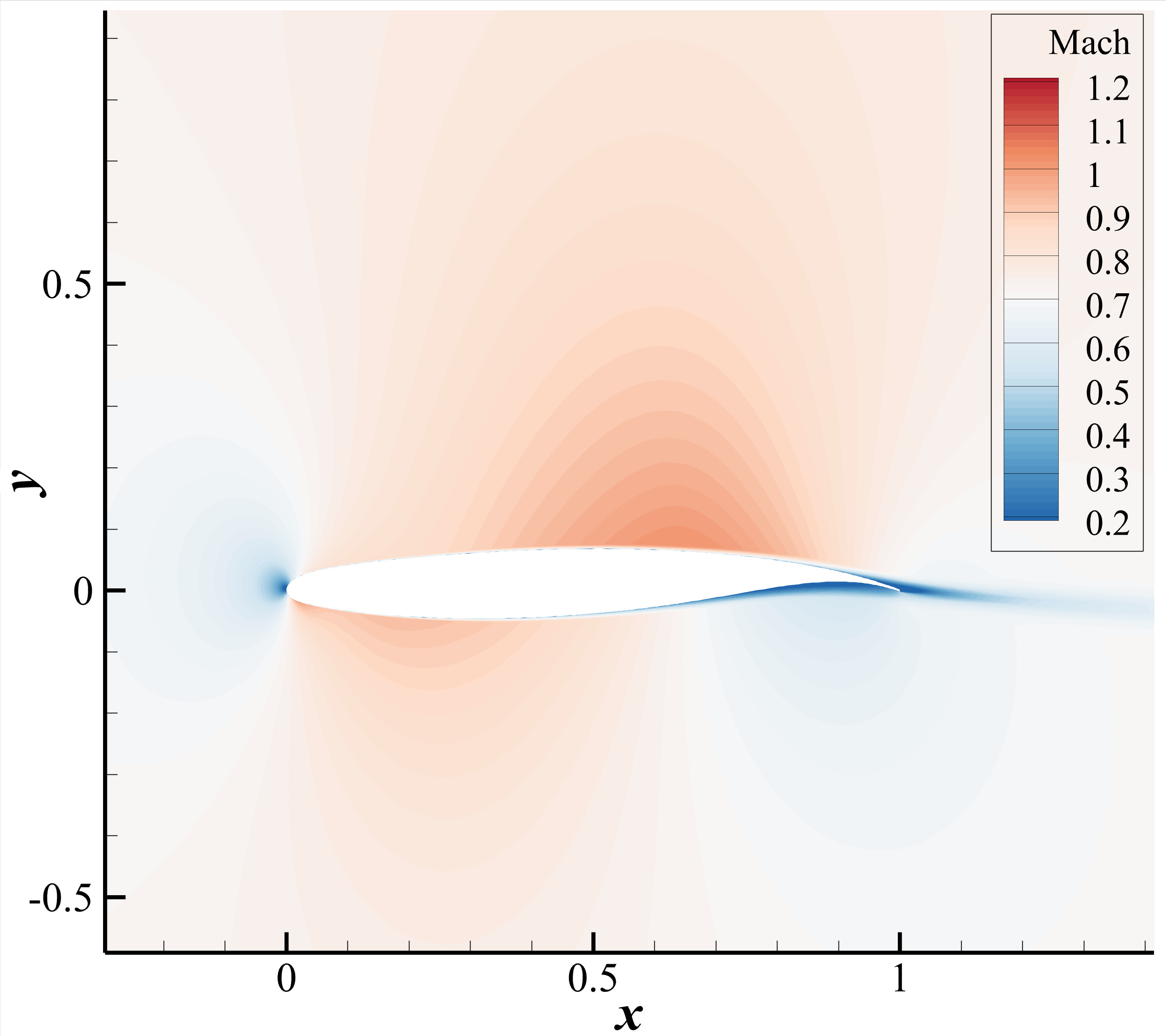}}
    \subfigure[$\alpha=-1.0^{\circ}$]{
    \includegraphics[width=0.3\textwidth]{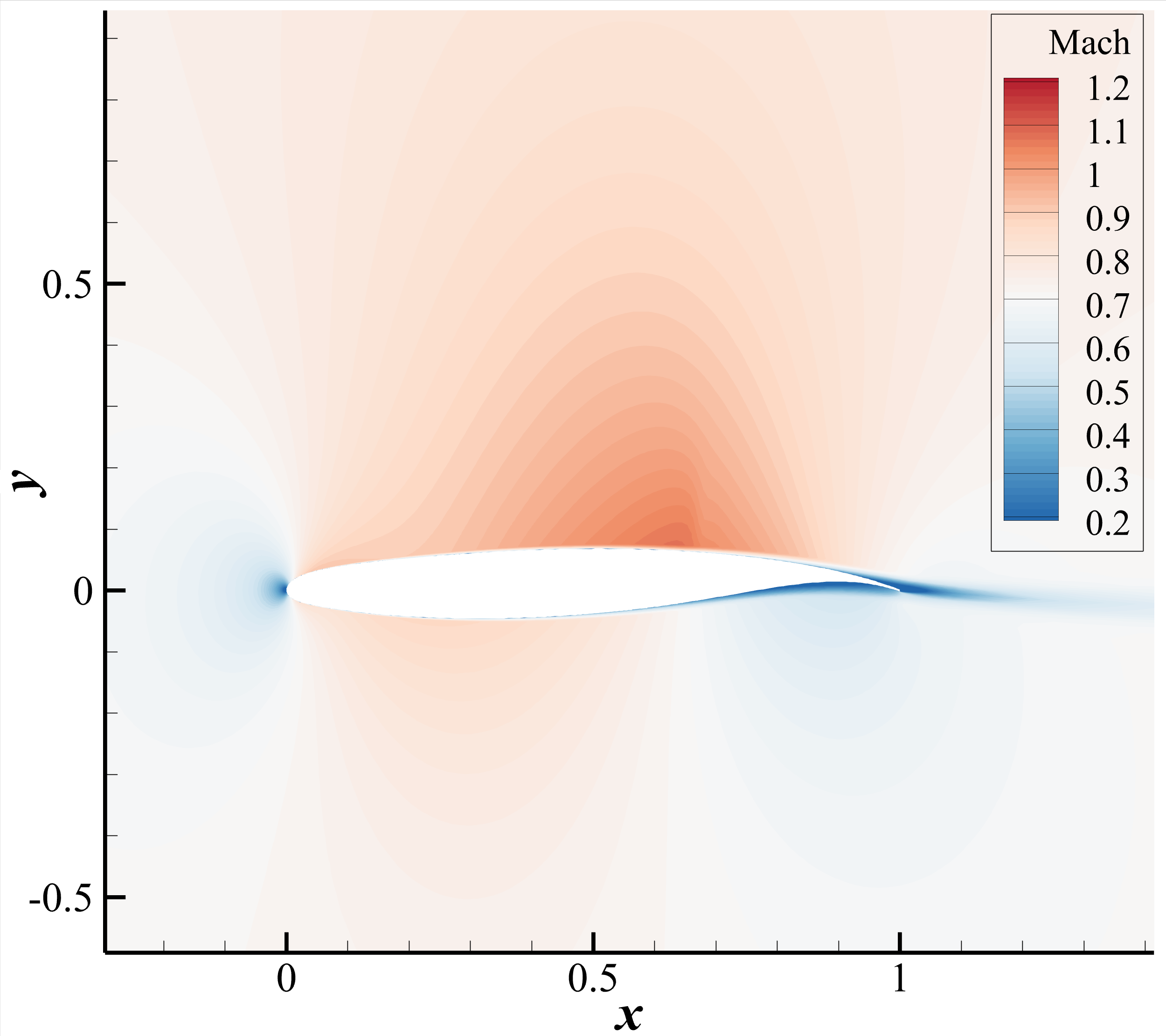}}
    \subfigure[$\alpha=0.0^{\circ}$]{
    \includegraphics[width=0.3\textwidth]{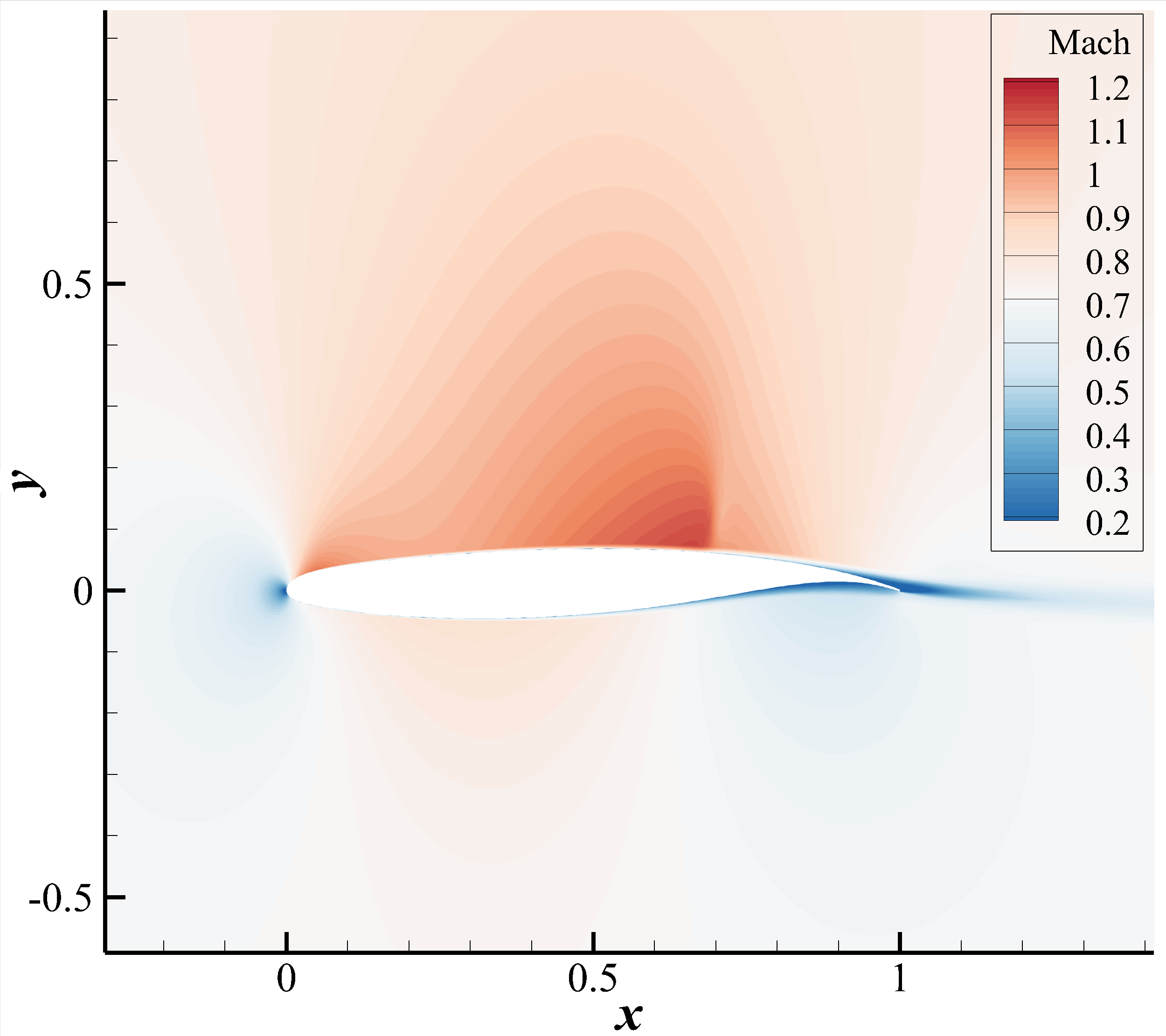}}
    \\
    \subfigure[$\alpha=1.0^{\circ}$]{
    \includegraphics[width=0.3\textwidth]{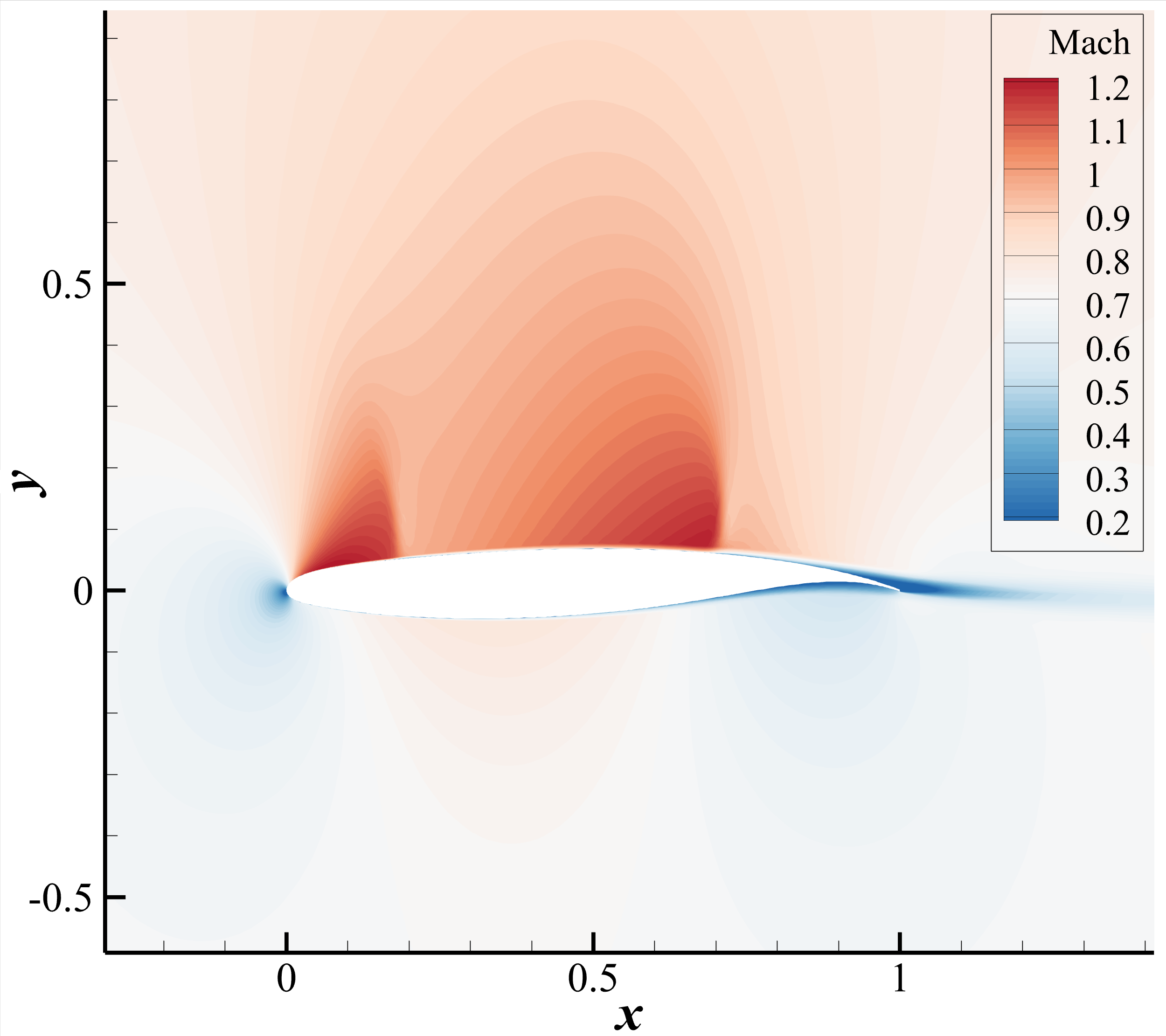}}
    \subfigure[$\alpha=2.0^{\circ}$]{
    \includegraphics[width=0.3\textwidth]{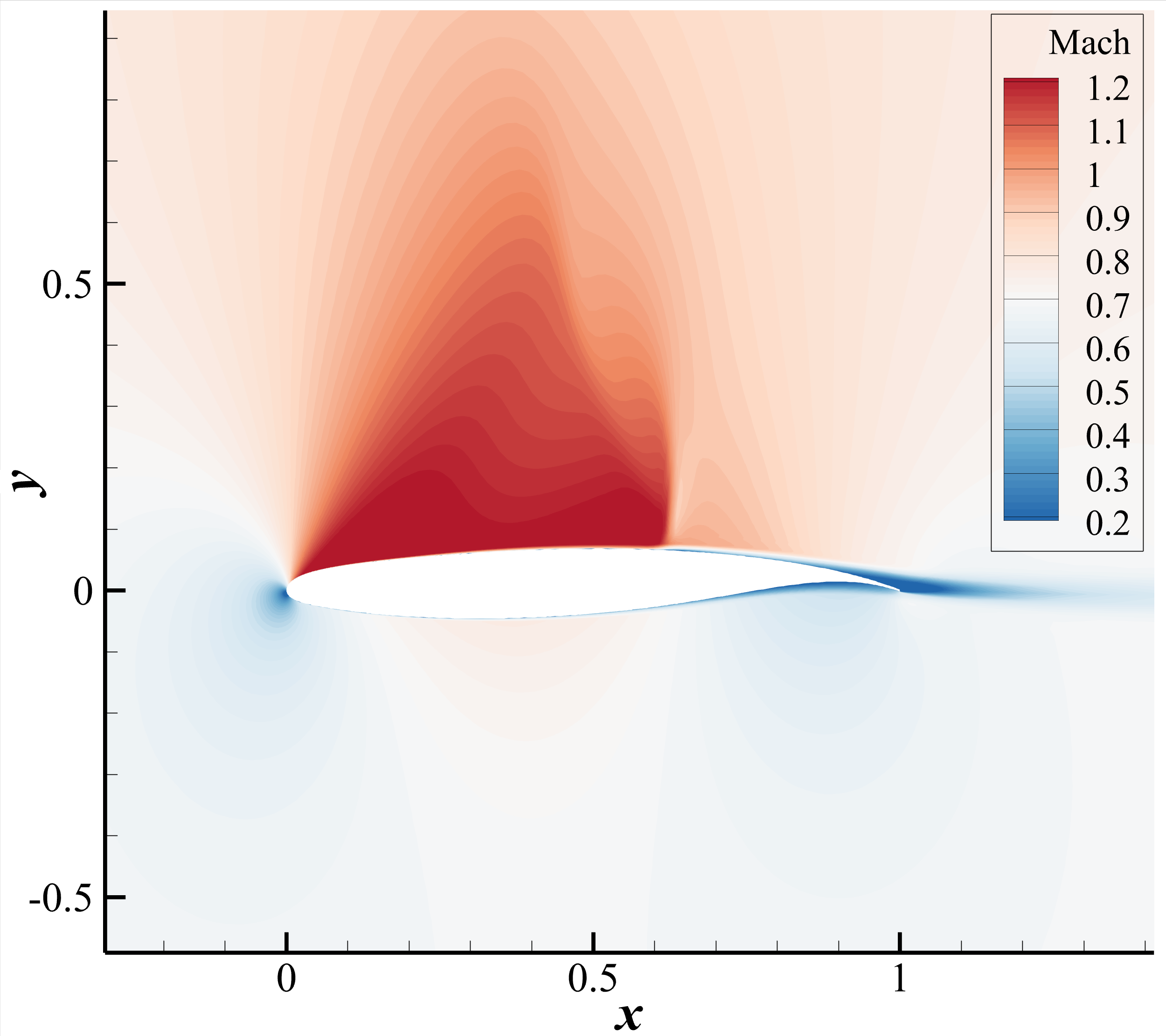}}
    \subfigure[$\alpha=3.0^{\circ}$]{
    \includegraphics[width=0.3\textwidth]{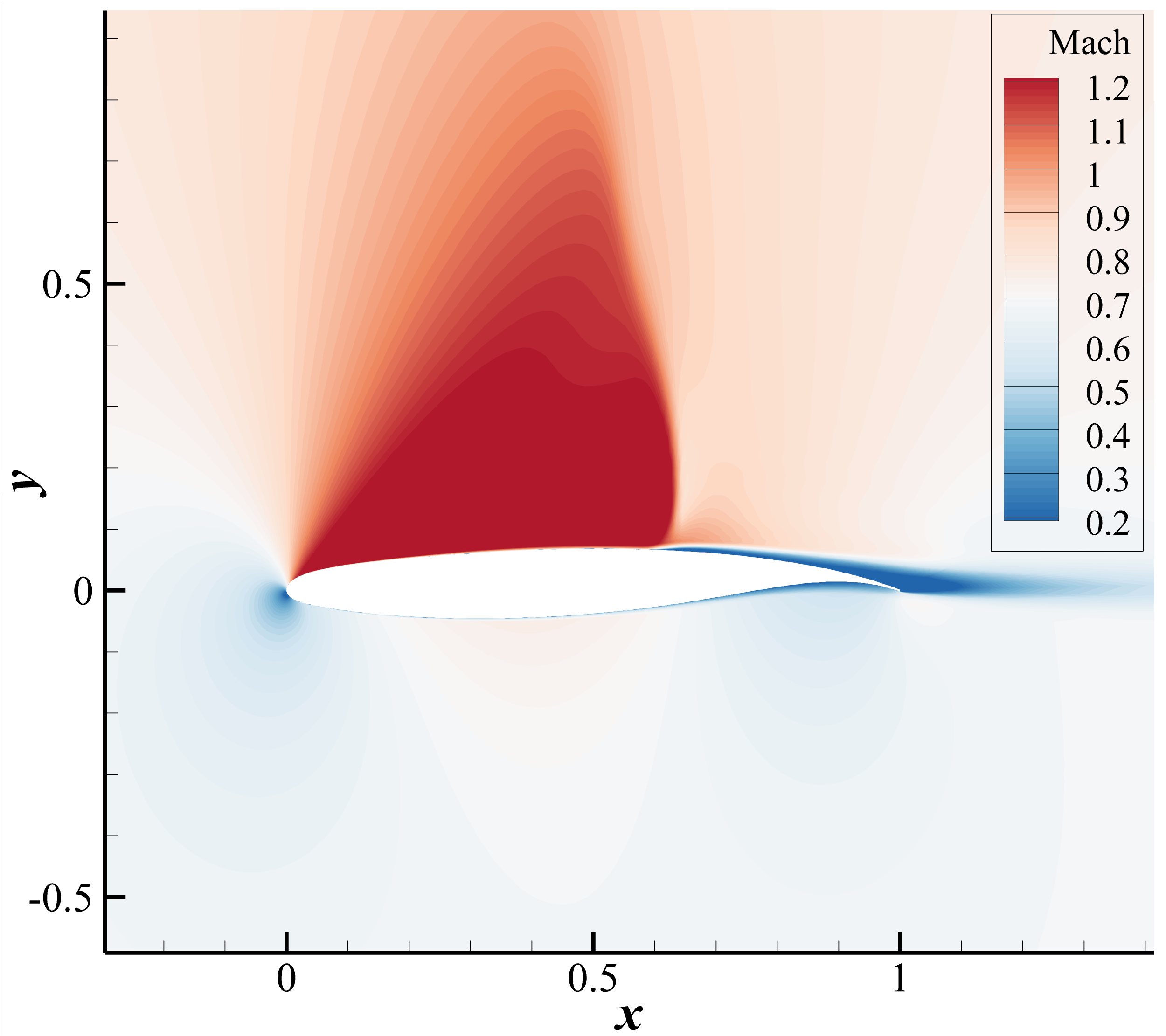}}
  \caption{Flow field for sample-g at different AoA.}
  \label{fig:fig16}
\end{figure}

In the above studies, sample-g exhibits relatively poor performance and we would like to further explore the reasons for this.  
The detailed flow fields of sample-g under various AoAs are presented in Fig.~\ref{fig:fig16}. 
When $AoA=-1^\circ$, a weak shock wave appears on the upper surface, and its intensity gradually strengthens as AoA increases. 
When $AoA=1^\circ$, a second shock wave appears on the upper surface near the leading edge. 
When $AoA=2^\circ$, two shock waves merge into a strong one as their respective ranges expand. 
The double shock wave is an undesirable flow phenomenon in engineering design, due to possible flow losses and poor performances at other flow conditions.  
On the one hand, the appearance of the first shock wave at small AoA indicates a compression effect, which can cause an early deviation from the linear assumption. 
On the other hand, the fusion process of the double shock wave is a typical nonlinear variation, leading to anomalous changes in the slope of the lift curve. 
As a result, it can be challenging for the model to achieve accurate prediction.

In this scenario, the shortcomings of the knowledge-embedded model are highlighted.
If the knowledge cannot fully represent the data, the knowledge-based network architecture may have limited performance at some outliers. 
This, in turn, can guide us to identify the key factors resulting in the poor performance, which can hardly attained in "black box" models.
As a result, the above findings will be conducive to developing the theories about the lift curve available to the double-shock wave patterns.
Moreover, we can analyse the characteristics of the airfoils susceptible to double shock waves and exclude these samples from the dataset.
Taking the Base Model as an example, if the testing samples with double shock wave are excluded (4.8\% of all testing samples),  MAE is $1.274\times10^{-3}$ decreasing by 16.0\% and MAXE is $3.002\times10^{-2}$ decreasing by 56.7\%. 
More significant improvement can be achieved if these undesirable samples are not involved in the training.


\section{Conclusion}
This paper proposes a knowledge-embedded meta-learning model to predict the lift coefficient of supercritical airfoil under various AoAs.
Three models, denoted as Base Model, +Weight Decay and -Activation, are designed so as to provide various interpretations.
Statistical results show that the Base Model achieves the highest accuracy with MAE as low as $1.517\times10^{-3}$, followed by -Activation and +Weight Decay. 
Comparable training and testing errors are achieved, indicating the excellent generalization performance of the model.
The primary network enables us to obtain several physical characteristics about the lift curve, such as $C_{L0}^{\prime}, C_{L\_crit}$.
Through the interpretable analysis, the influence of a certain airfoil geometry to them can be evaluated.
For example, the latter portion of the lower surface has the greatest positive influence on $C_{L0}$, which can help the designer to modify the shape in a right way.

More importantly, the research in this paper is an attempt to embed classical theoretical knowledge into the design of model architectures. 
In the field of aerodynamic design and even wider engineering, there are some theories that have been accumulated and validated for decades or even centuries.
Despite some of them are not universal, they can be extremely applicable and effective in certain situations.
The deep fusion of artificial intelligence algorithms with these domain knowledge can further enhance expressiveness and reliability, and also provides a possible direction for future collaborative innovation between human and machine.

Nevertheless, the proposed method still possesses limitations that require further investigation in the future.
Knowledge can be a restriction in some cases, although it may help construct stronger models in most circumstances.
Therefore, how to use knowledge more reasonably and even discover universal knowledge through the data is worthy of further study.


\bibliographystyle{elsarticle-num}
\bibliography{references}

\end{document}